%% file: mnras_template.tex
\documentclass[fleqn,usenatbib]{mnras}
\usepackage{float}
\usepackage{newtxtext,newtxmath}

\usepackage[T1]{fontenc}

\DeclareRobustCommand{\VAN}[3]{#2}
\let\VANthebibliography\thebibliography
\def\thebibliography{\DeclareRobustCommand{\VAN}[3]{##3}\VANthebibliography}

\usepackage{graphicx}	
\usepackage{amsmath}	
\usepackage{booktabs}  
\usepackage{multirow} 

\title[RedDots: GJ729 Magnetic field and companions]{RedDots: Magnetic field of the nearby active M dwarf GJ 729, and a search for companions}

\author[E. L. Brown et al.]{
E. L. Brown,$^{1,2}$\thanks{E-mail: e.brown@cqu.edu.au}
S. V. Jeffers,$^{3}$
S. C. Marsden,$^{2}$
F. Liebing,$^{4}$
S. Bellotti,$^{5,6}$
J.R. Barnes,$^{7}$
C.A. Haswell,${^7}$
\newauthor
and V. Koseleva${^8}$
\\
$^{1}$Central Queensland University, Cairns, QLD, 4870, Australia\\
$^{2}$Centre for Astrophysics, University of Southern Queensland, Toowoomba, QLD, 4350, Australia\\
$^3$Th\"uringer Landessternwarte Tautenburg, Sternwarte 5, 07778 Tautenburg, Germany\\
$^{4}$Private Astronomer\\
$^5$Leiden Observatory, Leiden University, PO Box 9513, 2300 RA Leiden, The Netherlands\\
$^6$IRAP, Universite de Toulouse, CNRS/UMR 5277, UPS-OMP, 14 Avenue E. Belin, Toulouse F-31400, France\\
$^7$Department of Physical Sciences, The Open University, Walton Hall, Milton Keynes MK7 6AA, UK\\
$^8$Institut für Astrophysik und Geophysik (IAG), Universität Göttingen, Friedrich-Hund-Platz 1, 37077 Göttingen, Germany}

\date{Accepted XXX. Received YYY; in original form ZZZ}

\pubyear{2026}

\begin{document}
\label{firstpage}
\pagerange{\pageref{firstpage}--\pageref{lastpage}}
\maketitle

\begin{abstract}
M dwarfs are prime targets for discovering exoplanets, and the nearest M dwarfs to the Sun provide among the best opportunities for follow-up detailed exoplanet characterization. GJ\,729, the seventh closest M dwarf to the Sun, presents significant challenges for exoplanet detection due to its high levels of magnetic activity. To address this, we present a detailed analysis of GJ\,729's magnetic field and its variability, followed by a search for exoplanets beneath the activity-induced noise in the stellar radial velocity. The geometry of GJ\,729's large-scale magnetic field was reconstructed using new and archival spectropolarimetric data for a total of four epochs spanning 10 years. Results indicate a weak large-scale field ranging from 50 to 145\,G, and an evolving non-axisymmetric field geometry that varies from poloidal dominated to a near-equal poloidal-toroidal configuration. We modeled activity-induced radial velocity variations using Gaussian Process Regression and activity diagnostics, and searched for planetary companions using $\sim90$\,d of high-cadence spectra taken contemporaneously with the high-precision CARMENES and HARPS spectrographs. Activity-only and activity + Keplerian models offered statistically equivalent fits, with a consistently preferred Keplerian period of $\sim7$\,d and amplitude of $\sim 1.9\rm{\,m\,s}^{-1}$ across a range of activity modeling approaches. This could relate to an Earth-mass or Super-Earth planet, or residual stellar activity with power concentrated at a multiple of the rotation half-period.  Our findings provide insight into the magnetic behavior of fully convective M dwarfs, and highlight the potential and challenges of detecting Keplerian RV signatures that are only a fraction of activity amplitudes.
\end{abstract}




\input{Intro}
\input{Data}
\input{ZDI}
\input{RV_Planet_Search}
\input{Discussion}

\section*{Acknowledgements}

We thank the anonymous referee for helpful comments that improved this work and the clarity of the paper. This work is based on observations obtained at the TBL, CFHT, the 3.6-m ESO telescope and the 3.5m telescope at Calar Alto Observatory. We thank the staff at each of these facilities for their time and data. We acknowledge use of the VALD database, operated at Uppsala University, the Institute of Astronomy RAS in Moscow, and the University of Vienna. S. Bellotti acknowledges funding from the project “Exo-space weather and contemporaneous signatures of star-planet interactions" (with project number OCENW.M.22.215 of the research programme “Open Competition Domain Science - M"), which is financed by the Dutch Research Council (NWO). J.R. Barnes and C.A. Haswell were funded by STFC under consolidated grant ST/X001164/1.

\section*{Data Availability}
 
Based on data from PolarBase\footnote{http://polarbase.irap.omp.eu}, the ESO archive\footnote{http://archive.eso.org/eso/eso\_archive\_main.html}, CARMENES data archive\footnote{http://carmenes.cab.inta-csic.es/gto/jsp/dr1Public.jsp} and MAST\footnote{https://mast.stsci.edu/portal/Mashup/Clients/Mast/Portal.html} archive at the Space Telescope Science Institute (STScI). 



\bibliographystyle{mnras}
\bibliography{bib} 




\appendix

\section{K2 photometry}\label{sec:K2phot}
\begin{figure*}
    \centering
    \includegraphics[width=0.9\linewidth]{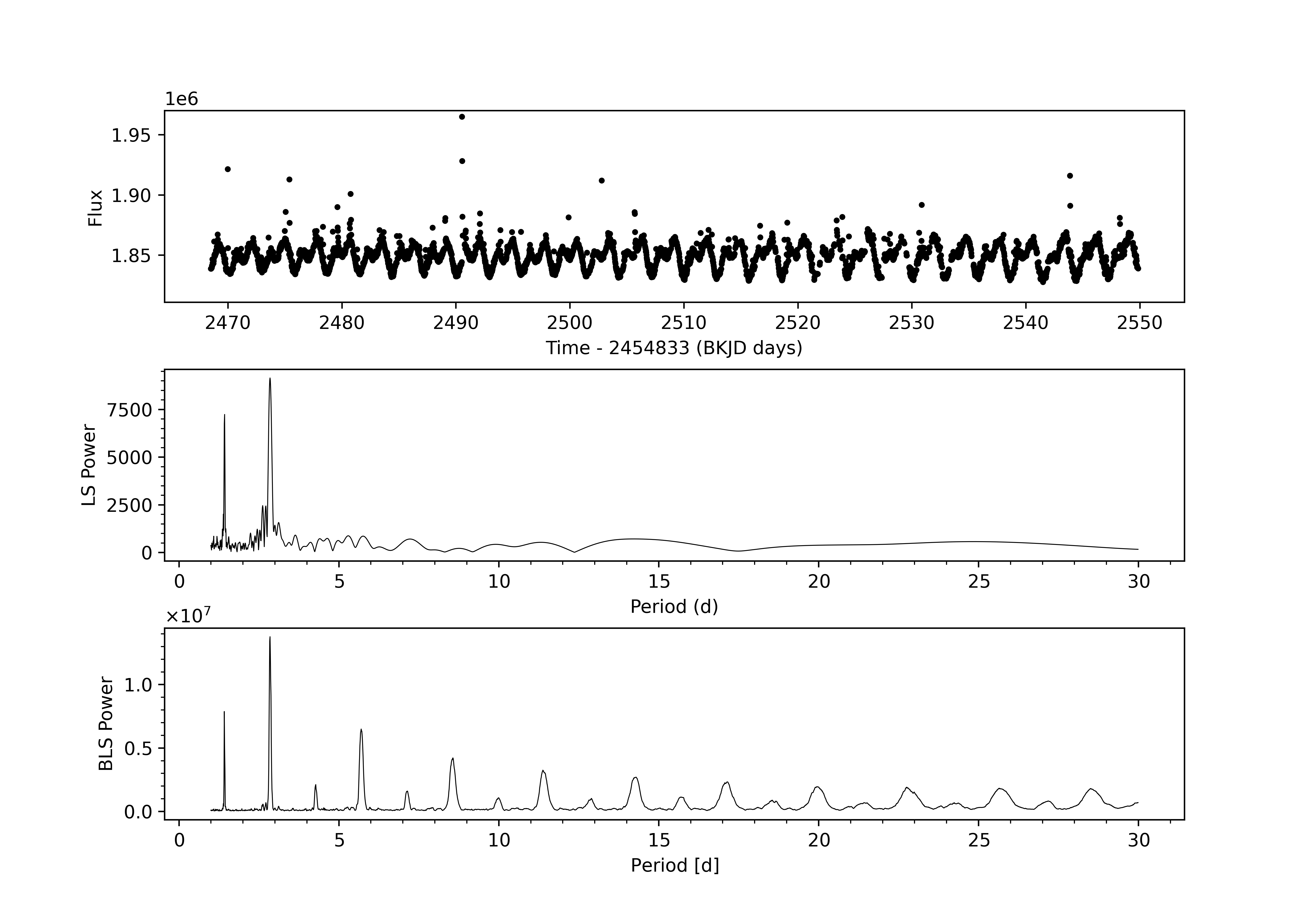}
    \caption{K2 photometric timeseries (top), Lomb-Scargle periodogram (middle) and Boxed Least-Squares periodogram (bottom) generated using Lightkurve \citep{Lightkurve}.}
    \label{fig:K2_timeseries}
\end{figure*}

K2 observations of GJ 729 (EPIC 215632123) from 2015 October 4 to 2016 April 22 show clear rotational modulation in Figure \ref{fig:K2_timeseries} with a period of $\sim2.848$d \citep{Ibanez2020} and a double dip pattern that prevails for the $\sim80$d of observations. The Lomb-Scargle periodogram (middle panel) shows that photometric power is concentrated at the rotation period and its first harmonic, with no significant sinusoidal signal near the candidate $\sim$7 d RV period. 

GJ 729 is not likely to host a detectable transiting planet \citep{Hardegree_Ullman_2023}. Nevertheless, we used a Boxed Least Squares (BLS) periodogram to search for transit-like signals. No significant transit-like feature was identified in the BLS periodogram (bottom panel of Figure \ref{fig:K2_timeseries}), including near the period of the candidate $\sim$7\,d RV signal. The BLS periodogram is dominated by stellar activity, with strong power at the stellar rotation period, its half-period, and a series of additional peaks at integer multiples of the rotation half-period. This is expected because the BLS algorithm fits box-shaped events to the repeating spot-induced flux minima, naturally producing peaks at integer multiples of the rotation half-period. One weak feature in this series of peaks lies near $\sim7$\,d, close to the period of the candidate RV signal. This demonstrates that stellar rotation can produce structured power at a range of related periods, which may also manifest in RV observations. However, unlike the photometry, the dominant rotational signal has been modelled and removed from the RVs (Figure \ref{fig:RV_fit}), leaving the $\sim7$\,d signal as the highest-probability residual sinusoidal signal above the remaining rotation-linked signals.

\section{ZDI results}\label{sec:ZDIresults_appendix}
\begin{figure*}
    \centering
    \includegraphics[width=0.9\linewidth]{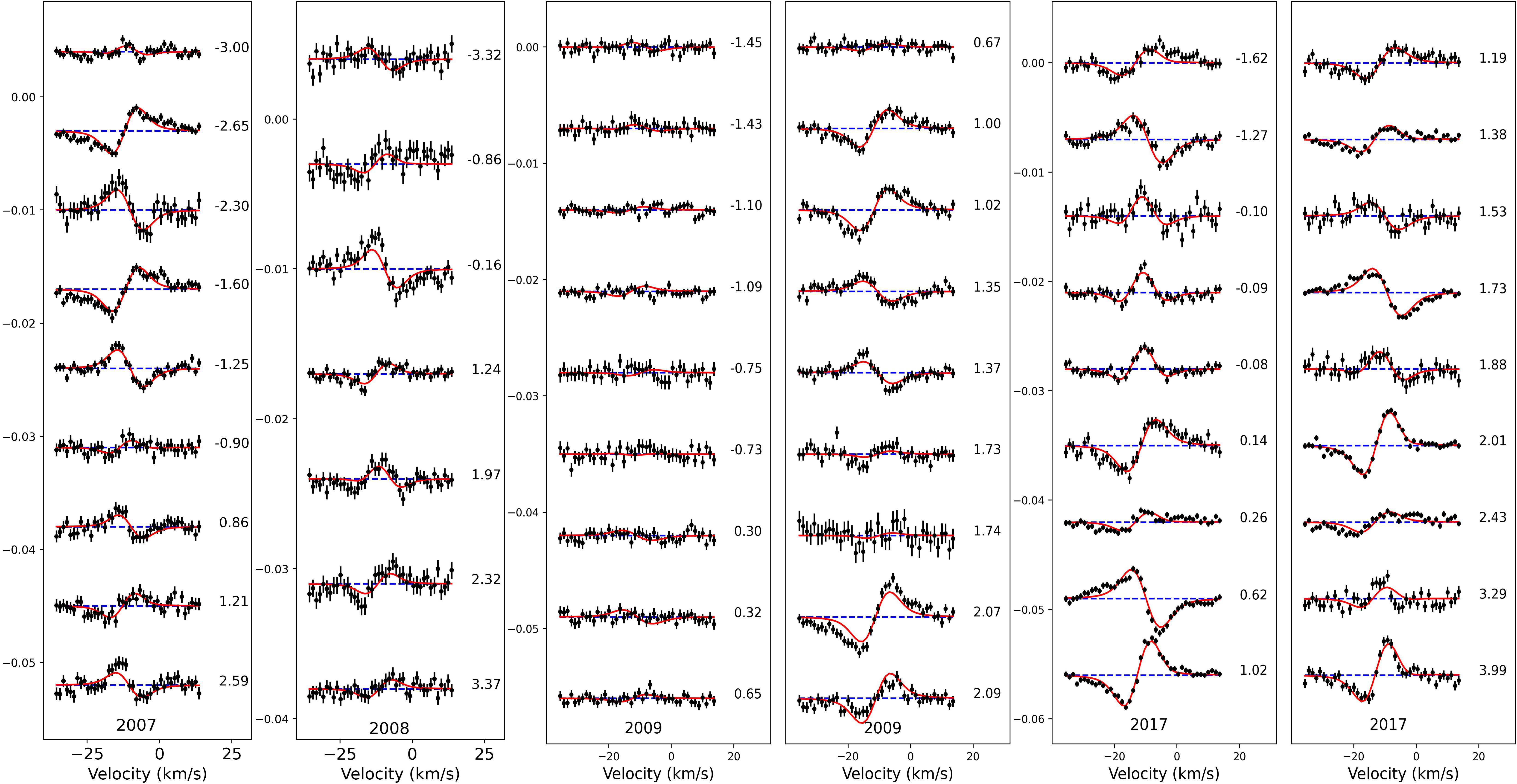}
    \caption{Model Stokes {\it{V}} profiles (red) and observed Stokes {\it{V}} LSD profiles (black) for 2007 July-August, 2008 August, 2009 August and 2017 August. Error bars shown are $\pm\,1\sigma$. The rotational cycles of each observation  with respect to the epochs in Table \ref{tab:observation_summary} are shown on the right of the profiles and assume $P_{rot}=2.848\textrm{\,d}$.}
    \label{fig:StokesV}
\end{figure*}

\begin{figure*}
    \centering
    \includegraphics[width=0.76\linewidth]{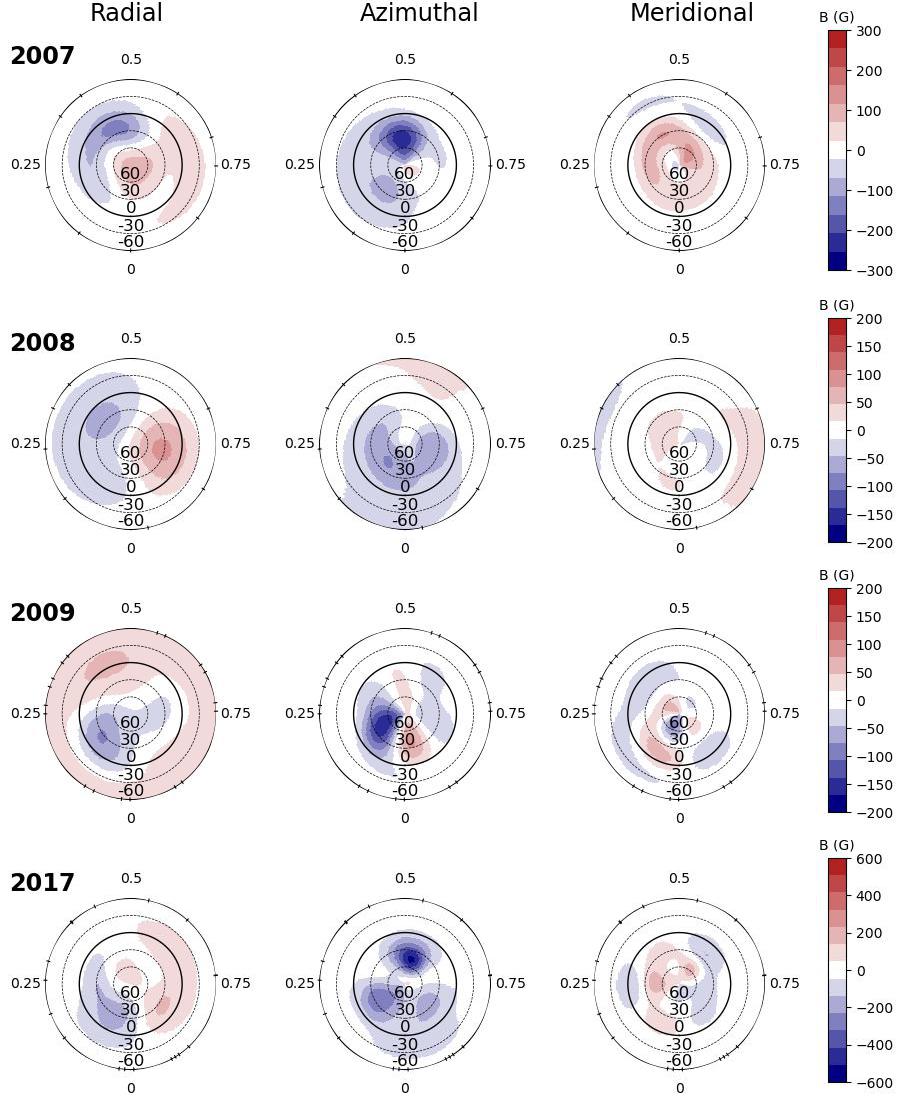}
    \caption{Reconstructed radial, azimuthal and meridional field maps for GJ729 from 2007 July-August through to 2017 August. The maps are flattened polar projections, showing the visible pole at the center and with latitudes extending down to -60$\deg$. Radial ticks indicate the phases of observations. Note that the colour scales differ for each epoch.}
    \label{fig:maps}
\end{figure*}

\begin{figure}
    \centering
    \includegraphics[width=\linewidth]{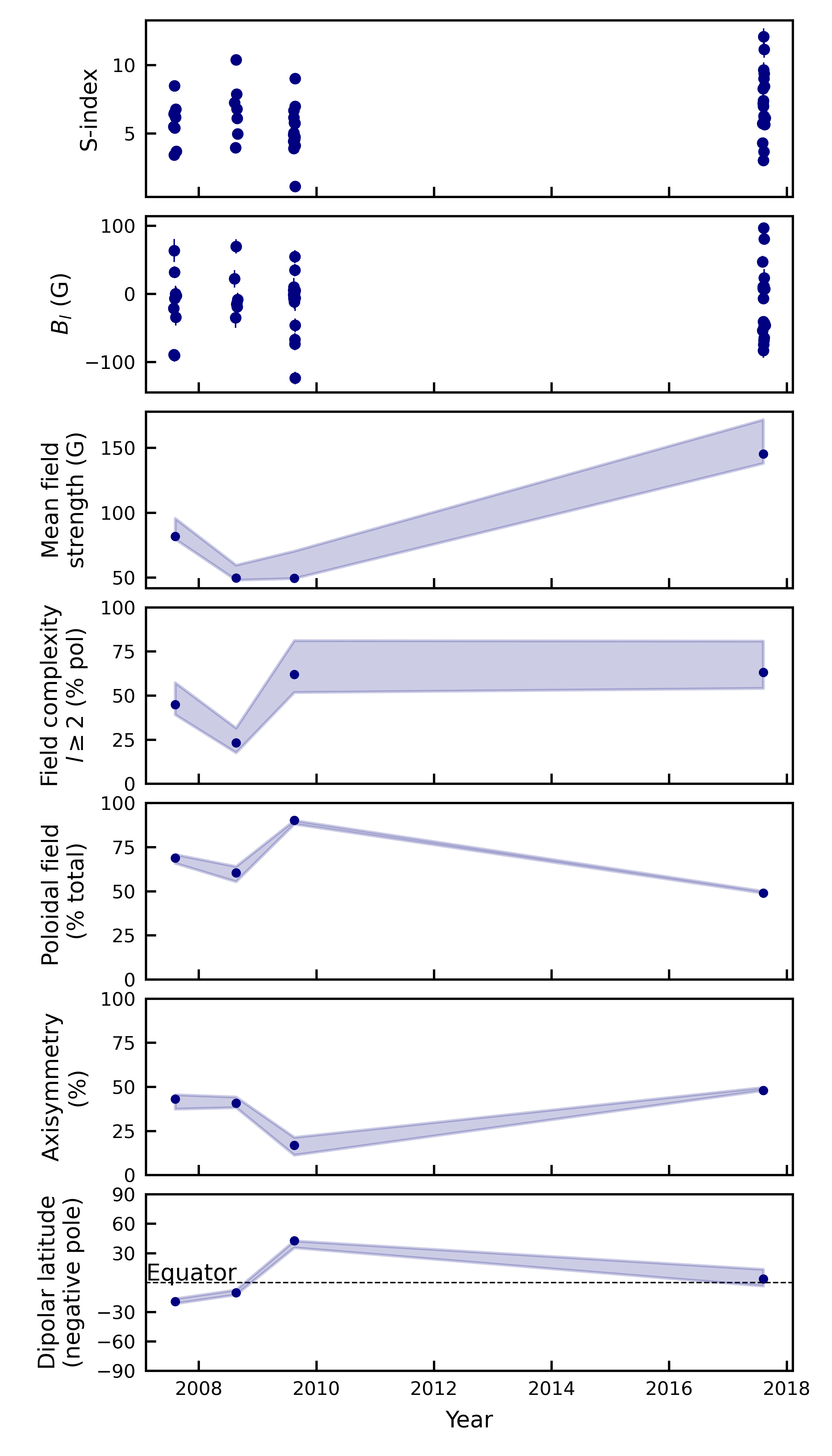}
    \caption{Time-series of the S-index and large-scale magnetic field properties as measured from NARVAL and ESPaDOnS data between 2007 and 2017. From top to bottom the axes show the Mount Wilson S-index, and surface-averaged longitudinal magnetic field strength ($B_l$), the mean magnetic field strength from ZDI, fraction of the poloidal field stored in quadrupolar and higher-order modes, poloidal field strength as a percentage of the total field strength, degree of axisymmetry of the field and the latitudinal location of the large-scale magnetic dipole (negative pole). The shaded regions indicate the sensitivity of our ZDI results to variations in the $v\sin i$ (2.7 - 3.3 $km\,s^{-1}$), inclination angle (37 - 70 $\degr$), rotation period (2.814 - 2.860 d) and rotational sheer (0 - 0.15 rad d$^{-1}$).}
    \label{fig:ZDI_properties_timeseries}
\end{figure}

Figures \ref{fig:StokesV} to \ref{fig:ZDI_properties_timeseries} show the observed and model Stokes {\it{V}} LSD line profiles, the large-scale magnetic field maps reconstructed using ZDI, and a timeseries of selected magnetic field parameters from Table \ref{tab:results}. 

\section{Mean-subtracted Stokes V profiles}

\begin{figure}
    \centering
    \includegraphics[width=0.45\columnwidth]{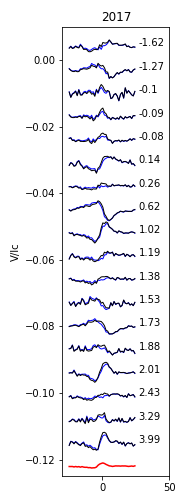}
    \caption{Stokes V observations from 2017 August (black), the time-averaged Stokes {\it{V}} profile (red) computed using all observations from 2007 to 2017, and the mean-subtracted profiles (blue). Rotational phases are shown to the right of the observations.}
    \label{fig:MeanStokesV2017}
\end{figure}

Figure \ref{fig:MeanStokesV2017} shows an example of the Stokes {\it{V}} profiles after subtracting the mean (or time-averaged) Stokes {\it{V}} profile. The mean Stokes {\it{V}} profile is small compared to the mean-subtracted Stokes {\it{V}} profiles.

\section{Carmenes and HARPS RVs}

\begin{figure*}
    \centering
    \includegraphics[width=\linewidth]{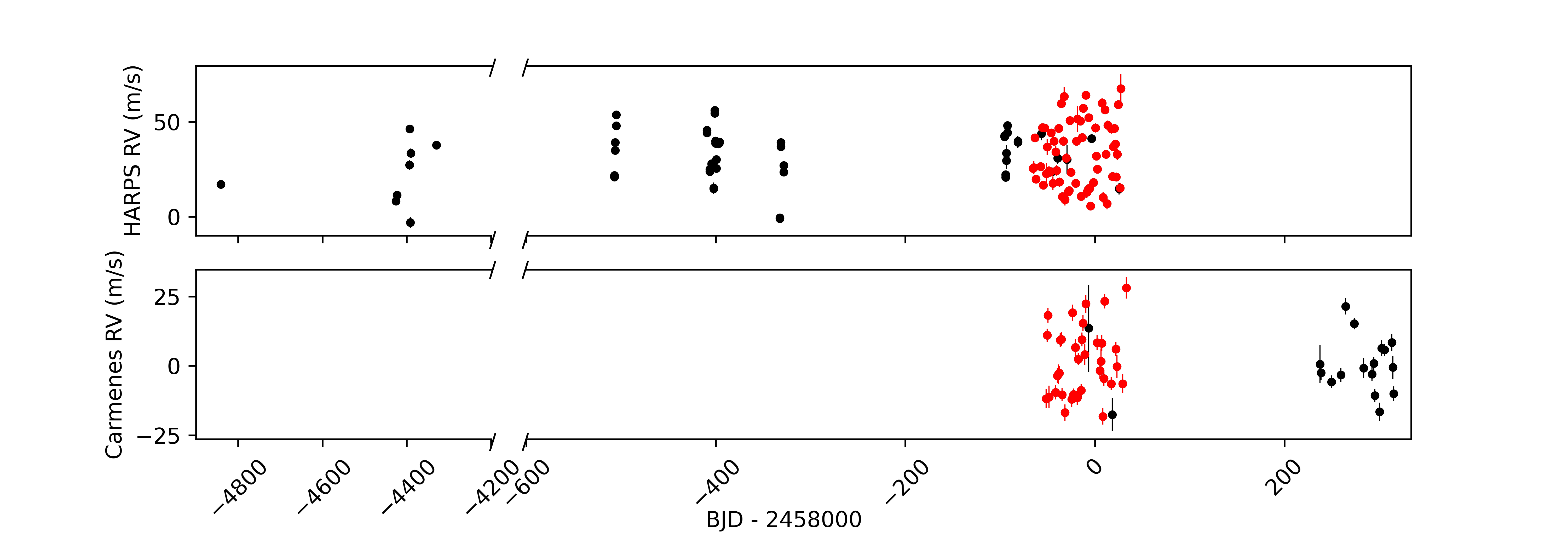}
    \caption{HARPS (top) and CARMENES (bottom) RVs. The data adopted in this study are shown in red. }
    \label{RVobs_timeseries}
\end{figure*}

Figure \ref{RVobs_timeseries} shows all HARPS and CARMENES RVs available to the public.

\section{HARPS S-index}

\begin{figure}
    \centering
    \includegraphics[width=\columnwidth]{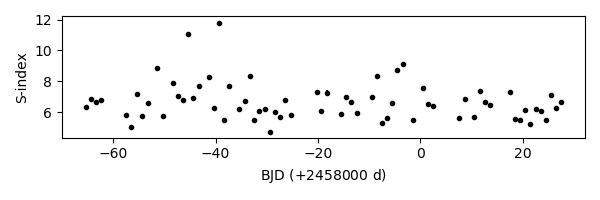}
    \caption{Timeseries of the Mount Wilson S-index measured using the doubly ionized Calcium H and K lines in the HARPS spectra. }
    \label{fig:HARPS_Sindex}
\end{figure}

Figure \ref{fig:HARPS_Sindex} shows the S-index (a measure of chromospheric activity) measured using HARPS observations of GJ 729. CARMENES does not cover the wavelength required to measure the S-index. 

\section{Effect of alternative activity models and data sets in PyORBIT}\label{sec:RVtests}
\begin{figure}
    \centering
    \includegraphics[width=0.9\linewidth]{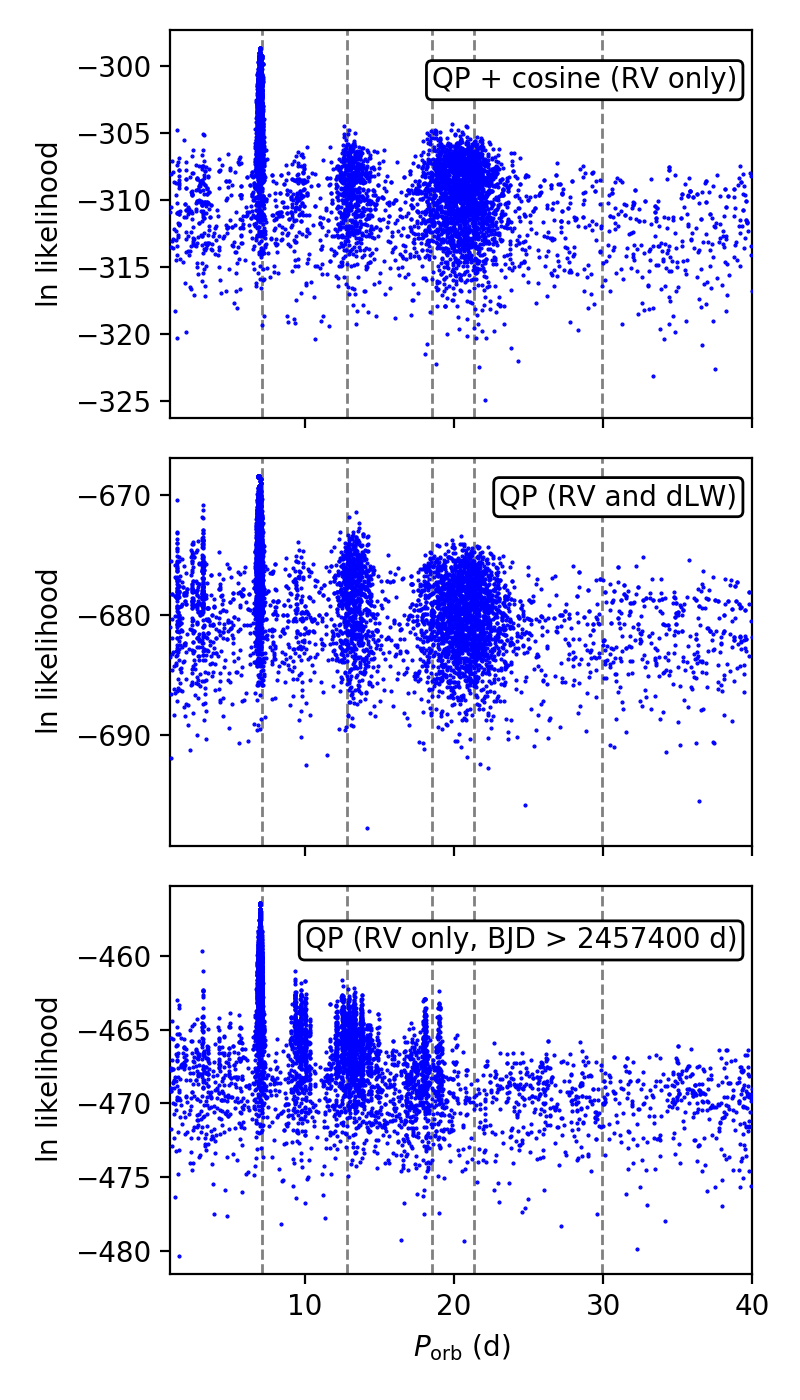}
    \caption{Same as Figure \ref{fig:PversusLnProb}, but for alternative activity models tested with {\sc{PyORBIT}}. Plots show the log-likelihood versus planet orbital period for each sample when (top) a quasi-periodic + cosine GP is fitted to the RV, (middle) a QP kernel is jointly fitted to the RV and dLW, and (bottom) a QP model is fitted to an extended RV timeseries. }
    \label{fig:RVtests}
\end{figure}

We tested the sensitivity of the $\sim7$\,d Keplerian recovered with {\sc{PyORBIT}} when using an alternative QP + cosine GP kernel for activity, when jointly fitting a QP GP to the RV and the dLW activity index, and when including additional HARPS and CARMENES data from BJD 2457500 d onward. Figure \ref{fig:RVtests} shows the log-likelihood as a function of orbital frequency for samples drawn from these models. The grey vertical lines indicate odd multiples of half the rotation period ($3 \times \frac{P_{rot}}{2}$,  $5 \times \frac{P_{rot}}{2}$ etc.).  Each model favours a Keplerian signal at $7$\,d over Keplerians at periods up to 40\,d.  The best-fitting planet parameters for each tested model are reported in Table \ref{tab:GP_results_tests}. The planet parameters are consistent across our tests, and in each case the GP + Keplerian model is statistically equivalent to a GP only model. 

\begin{table*}
    \centering
    \caption{Best-fitting Keplerian parameters for different activity models tested with {\sc PyORBIT}. Priors on orbital parameters, and the results for QP (RV only) are the same as in Table \ref{tab:GP_results}. }
    \begin{tabular}{lcccc}
    \toprule
     \bf{Parameter} & \bf{QP (RV only)} & \bf{QP + cosine (RV only)} & \bf{QP (RV + dLW)} & \bf{QP (RV only, extended dataset)} \\ \midrule
    $P_{\rm{orb}}$ (d) & $7.11\pm0.84$ [6.99] & $7.04\pm0.45$ [6.95] & $6.99\pm0.56$ [7.02] & $12.73\pm7.74$ [7.00] \\
    $K$ (m\,s$^{-1}$)  & $1.61\pm0.64$ [1.88] & $1.85\pm0.56$ [2.29] & $1.75\pm0.62$ [2.15] & $1.49\pm0.81$ [1.86] \\
    $T_c$ (BJD - 2458000 d) & $5.45\pm2.37$ [6.58] & $6.05\pm1.99$ [6.66] & $2.89\pm3.38$ [6.76] & $6.02\pm1.80$ [6.86] \\
    \midrule
    Bayes Factor, $\Delta \ln Z$ & $+1$ & $+2$ & $+1$ & $+0$\\
    \bottomrule
    \end{tabular}
    \label{tab:GP_results_tests}
\end{table*}

\section{GP + Keplerian model parameters}

\begin{figure*}
    \centering
    \includegraphics[width=\linewidth]{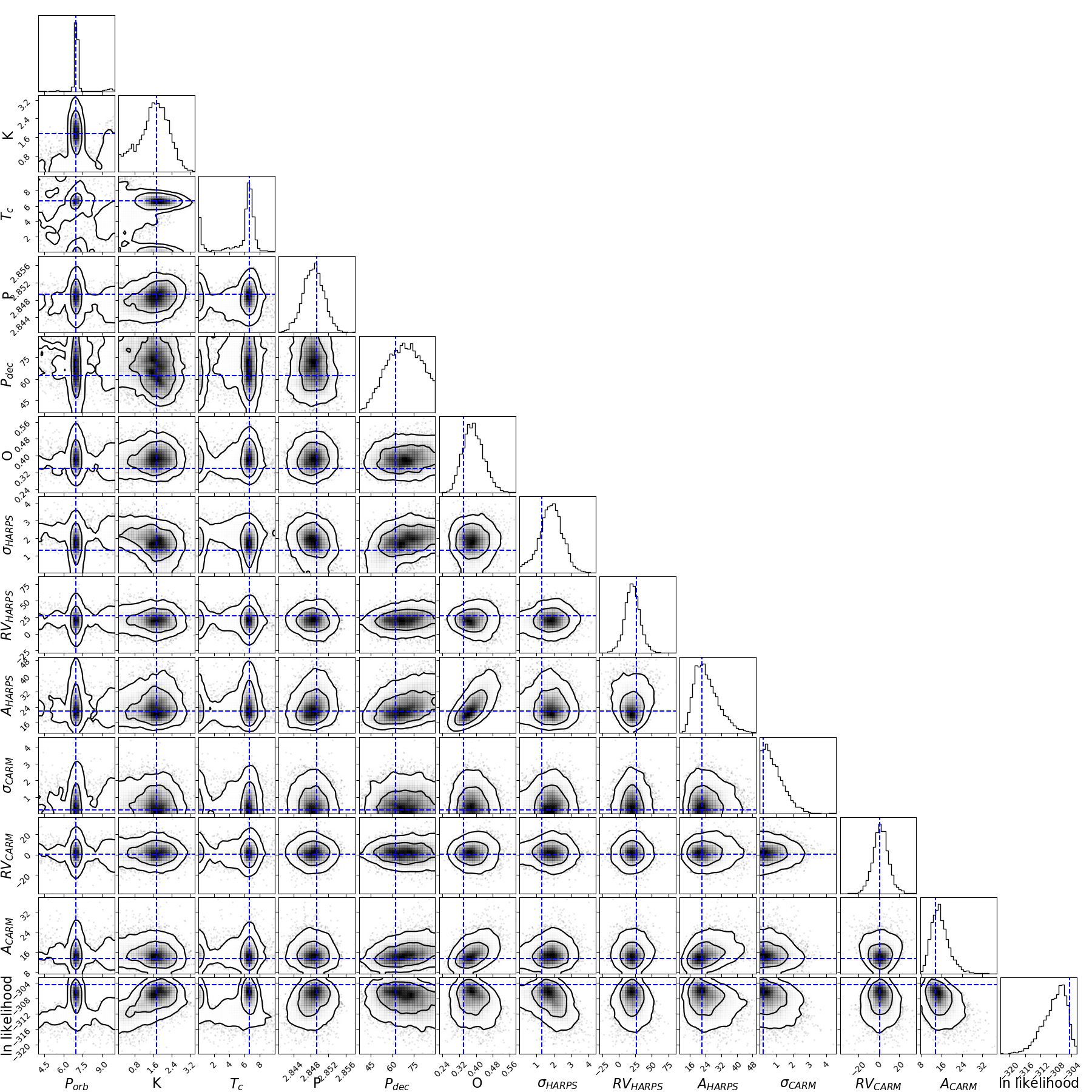}
    \caption{Posterior distribution of the fitted parameters for a GP + Keplerian model to the HARPS and CARMENES RVs with PyORBIT. Contours show the $1\sigma$ and $2\sigma$ posterior distributions, and the dashed blue lines indicate the maximum-a-priori parameters.}
    \label{fig:corner_PYORBIT}
\end{figure*}

The corner plot in Figure \ref{fig:corner_PYORBIT} shows the parameter correlations and posterior distributions for the GP + Keplerian model fitted to RVs with {\sc{PyORBIT}} .

\bsp	
\label{lastpage}
\end{document}

%% file: Intro.tex
\section{Introduction}

M dwarfs are the most common stars in our Galaxy and comprise around two thirds of the stellar population in the solar neighborhood \citep{Davy_Kirkpatrick_2012}. They make attractive targets in the search for exoplanets orbiting within the habitable zone (HZ), commonly defined as the circumstellar region wherein liquid water can be sustained on a planet's surface. The low luminosities of M dwarfs mean that HZ planets orbit close to the star, with short orbital periods, making them accessible for detection in short observational campaigns. The low masses of these stars means that even Earth-sized planets could impart potentially detectable radial velocity (RV) amplitudes (\citealt{Jeffers2020_RedDots}), enabling their discovery regardless of whether the planet transits the host star. 

However for magnetically active M dwarfs, activity can complicate the detection, characterization and potential habitability of exoplanets \citep{Vidotto_2013,Cameron2020}. Star spots and bright plages introduce spectral line asymmetries that shift line centers, inducing spurious RV variations that can obscure or even mimic true stellar RV 3variations imparted by exoplanets \citep{Barnes2011,Jeffers2014b,Nava2020}. Magnetic activity is driven by the stellar magnetic field, which is closely linked to stellar structure and evolution. Monitoring the magnetic fields of M dwarfs is therefore crucial not only for addressing challenges in exoplanet detection and refining habitability criteria beyond the presence of liquid water, but also for insight into the shared evolution of star-planet systems.

In this paper we conduct a combined search for exoplanets and a study of the large-scale magnetic field evolution of the active M dwarf GJ 729, the 7th nearest M dwarf to the Sun\footnote{www.recons.org}. We search for planetary companions using precise RV measurements from the RedDots \citep{Jeffers2020_RedDots} and CARMENES (Calar Alto high-Resolution search for M dwarfs with Exoearths with Near-infrared and optical Échelle Spectrographs) programs. Zeeman Doppler Imaging (ZDI) and Principle Component Analysis (PCA) are used to study the evolution of the large-scale magnetic field of GJ 729 between 2007 and 2017, and inform our planet search. The work represents a valuable test case for detecting planets around magnetically active stars. 

The remainder of the paper is organised as follows; in Section \ref{sec:MFoMD} we describe the magnetic fields and activity of M dwarfs. Sections \ref{sec:GJ729} and \ref{sec:data} describe GJ 729 and the data used in this study. In Section \ref{sec:magmapping} we describe the ZDI and PCA techniques, and the long-term evolution of the magnetic field of GJ 729. Section \ref{sec:RVs} describes RV extraction, activity modelling and the search for exoplanets. Key findings are summarized in Section \ref{sec:D&C}. 

\section{Magnetic fields of M dwarfs}\label{sec:MFoMD}

M dwarfs straddle the transition between partially convective and fully convective stellar structures, thought to take place around 0.20 to 0.35$M_{\odot}$ \citep{Chabrier1997, Dorman_1989, Rabus2019}. Early-type M dwarfs ($\sim0.35-0.6M_{\odot}$) have a radiative core and a convective envelope, like more massive solar-type stars, while late-type M dwarfs  ($\lessapprox0.35M_{\odot}$) are fully convective. It is not yet clear how this structural distinction influences their magnetic dynamos. 

In the Sun and other stars with both convective and radiative zones, the cyclic transformation and re-generation of the large-scale magnetic field is thought to be driven by an $\alpha$-$\Omega$ dynamo seated at the tachocline, the interface between the radiative and convective zones \citep{Charbonneau2005}. Rotational shear at the tachocline coupled with differential rotation between the stellar equator and poles are thought to convert a predominantly dipolar, axisymmetric (i.e. the magnetic dipole is aligned with the stellar rotational axis) and poloidal field to a slightly more complex, non-axisymmetric, toroidal structure (the $\Omega$ effect). The predominantly poloidal field is then regenerated through cyclonic convection or the Babcock-Leighton mechanism (the $\alpha$ effect). This dynamo manifests at the stellar surface in activity cycles, observable in the Sun through e.g. the spot cycle, or in more distant stars through cyclically varying photospheric brightness and/or chromospheric emissions \citep{Baliunas1995, Radick1998}. 

For fully-convective stars, Zeeman Doppler Imaging (ZDI, \citealt{Semel1989, Donati1997b}) reconstructions of global-scale magnetic fields have shown either strong (kG), stable, predominantly dipolar and axisymmetric fields, or weaker (a few hundred G), multipolar and non-axisymmetric fields that evolve over year(s) and can be dominantly toroidal or poloidal \citep{Morin2008a,Morin2010,Kochukhov2017,Bellotti2024}. The two types of magnetic fields observed across mid-late M dwarfs could indicate a dynamo bi-stability \citep{Morin2011} (i.e. there are two distinct dynamos operating in these stars) or a cyclic dynamo with similar underlying mechanisms to the solar dynamo, but where the tachocline plays a lesser role  \citep{Kitchatinov2014,Wright_2016, Wright_2018}. 

Solar-like magnetic cycles have not yet been observed for any M stars, but a large-scale polarity reversal was detected for the fully convective, slowly rotating ($\rm{P_{rot}\sim176}$ d), M4.5V star GJ 1151 \citep{Lehmann2024}. Large-scale magnetic polarity reversals are an important component of the solar magnetic cycle and those of several other F-K stars \citep{Mengel2016, BoroSaikia2018_61cygA, Brown2021, Lehtinen2022, Willamo2022_5SolarTypeStars}.  Additionally, possible activity cycles have been detected in chromospheric and photospheric observations for M stars \citep{SuarezMascareno2016,Ibanez2020,Fuhrmeister2023}. A variety of secular magnetic field trends have been observed in the longitudinal magnetic field strength, from stable to pulsating or sinusoidal trends \citep{Bellotti2024}. An $\alpha^2$ dynamo could operate in these stars, where turbulent motions within the convective envelope produce irregular magnetic field variations, without the need for a tachocline. Continued, long-term monitoring is essential to clarify these behaviors and better understand the magnetic dynamos operating within M dwarfs.

\section{GJ 729}\label{sec:GJ729}

GJ 729 (Ross 154, V* V1216 Sgr, CD-23 14742, HIP 92403) is an M3.5 dwarf with the parameters shown in Table \ref{tab:stellar_params}. GJ\,729 is the 7th nearest M dwarf to the Sun, making it an excellent candidate for detailed exoplanet characterization in the future. It is a known flare star, with 
high-amplitude activity-induced RV variability that makes exoplanet detection challenging. Previous searches for sub-stellar companions and circumstellar material around GJ 729 have yielded non-detections. No brown dwarf or massive Jupiter companion was found by \citet{Schroeder2000} or \citet{Tanner2010}, and \citet{Plavchan2005} found no evidence of a debris disk at 11.7$\mu$m. \citet{Hardegree_Ullman_2023} estimated a habitable zone for GJ729 between a semi-major axis of 0.064-0.127 au. Assuming a 1 $R_{\oplus}$ planet in the middle of the star's habitable zone, they determined an orbital period of 25.68 d, with a geometric transit probability of only 1.07 percent.  

GJ 729 is an interesting target in terms of its magnetic behaviour, being a fully-convective star \citep{Irving_2023} that shows evidence for long-term activity cycles, without any differential rotation \citep{Ibanez2020}, contrary to a solar-type dynamo.  \citet{SuarezMascareno2016} found two coexisting photometric cycles with lengths of 7.1 $\pm$ 0.1 and 2.1 $\pm$ 0.1\,yr using 8.6\,yr of photometric observations from the All Sky Automated Survey (ASAS). Reanalysis of the data by \citet{Irving_2023} with new observations from ASAS-SN indicated a $\sim3.5\pm1.1$\,yr cycle. \citet{Ibanez2020} detected a 4.2\,yr activity cycle in the chromospheric S-index, possibly modulated by a much longer activity cycle that potentially reached a grand minimum around 1998 to 2004. 
\citet{Ibanez2020} found that the two cycles are roughly consistent with the active and inactive branches in the rotation period - activity cycle period space \citep{Bohm2007}.  They also reconstructed spot models using K2 photometry, finding that spots do not migrate between active longitudes, indicating that the activity cycles are not solar like, probably driven by an $\alpha^2$ dynamo. 

\begin{table}
    \centering
    \caption{Stellar parameters for GJ 729 }
    \begin{tabular}{lcc}
    \toprule
    \textbf{Parameter} & & \textbf{Reference} \\
    \midrule
        Spectral type   &   M3.5V & \\
        Distance & 2.9762 pc & \citet{Gaia_distances} \\
        Age & 0.2 - 0.6 Gyr & \citet{Ibanez2020} \\
        $M$ & $0.184\pm0.011$ $M_{\odot}$ & \citet{Schweitzer2019} \\
       $R$ & $0.197\pm0.006$ $R_{\odot}$ & \citet{Schweitzer2019} \\
        ${T_{\rm{eff}}}$ & $3340\pm30$ K & \citet{Marfil2021} \\
        $\log g$ & $5.23\pm0.11$  &  \citet{Marfil2021} \\
        Fe/H & $-0.63\pm0.1$ & \citet{Marfil2021} \\
        $\log R'_{HK}$ & $-4.405\pm0.510$ & \citet{Engle_2023} \\
        $\log(\langle L_X \rangle / L_{Bol})$ & $-3.28\pm0.24$ & \citet{Engle_2023} \\
        $v\sin{i}$  & $3.0\pm1.5$ km\,s$^{-1}$ & \citet{Reiners2018} \\
        Inclination angle & $37\pm18$\degr  & \citet{Reiners2018} \\
        ${P_{\rm{rot}}}$ & $2.848\pm0.001$ d & \citet{Ibanez2020} \\
        \bottomrule
    \end{tabular}
    \label{tab:stellar_params}
\end{table}

%% file: Data.tex
\section{Data}\label{sec:data}

\subsection{High-precision spectra from HARPS and CARMENES}

As part of the RedDots campaign \citep{Jeffers2020_RedDots} GJ 729 was observed with the High Accuracy Radial velocity Planet Searcher (HARPS) spectrometer at the 3.6-m ESO telescope (La Silla Observatory, Chile, \citealt{Mayor2003}) roughly nightly for $\sim3$ months from 2017 July to September, for a total of 68 observations (Program ID 099.C-0880, PI: Anglada-Escude). After removing outliers (see Section \ref{sec:RVs} for details) the data totaled 62 observations. A further 47 archival HARPS observations were available,  taken in 2004-05 (ID: 072.C-0488, PI: Mayor) and 2016-17 (ID: 097.C-0864, PI: Lannier), but we excluded these from our planet search due to the large phase gaps and the HARPS fiber upgrade that occurred in May 2015 (BJD = 2457172, \citealt{LoCurto_2015}).  HARPS spectra have a resolution of $\sim$110000 and spectral coverage from 3800 to 6900 {\AA}, with an RV precision down to 1\,m\,s$^{-1}$ \citep{Trifonov2020}. The HARPS data were downloaded from the ESO archive in the reduced form (DRS v3.5, \citealt{Trifonov2020}).

GJ 729 (Karmn J18498-238) was also observed as part of The CARMENES search for exoplanets around M dwarfs \citep{CarmenesProject,Reiners2018} using the CARMENES spectrograph \citep{CarmenesInstrument,CarmenesInstrument2} at the 3.5m telescope at Calar Alto Observatory, Spain, over 40 nights between 2017 July and October with a cadence of once every 1-10\,d. We removed 2 data points based on our outlier analysis in Section \ref{sec:RVs}. A further 16 observations were taken between April and October 2018, which we also excluded from our RV planet search due to the large phase gap. We make use exclusively of the publicly available Data Release 1 (DR1) observations, which correspond to spectra obtained with the visible (VIS) channel. The VIS channel has a wavelength range of 5200 to 9600{\AA}, spectral resolution of 80000 to 100000 and radial-velocity accuracy of $\sim1\rm{m s^{-1}}$ with long-term stability \citep{Quirrenbach2020}. Reduced and calibrated spectra, along with radial velocities measured using SERVAL, are available in the CARMENES GTO Data Archive, DR1 \citep{Ribas2023}. 

\subsection{NARVAL and ESPaDOnS spectropolarimetric observations}

To study the large-scale magnetic field of GJ 729 we used 52 circularly polarized (Stokes {\it{V}}) spectra available to the public on PolarBase\footnote{http://polarbase.irap.omp.eu} \citep{Petit2014} in their reduced and calibrated state. The data included 42 observations from four separate observing runs in 2007 July-August, 2008 August, 2009 August and 2017 August with the NARVAL\footnote{http://www.ast.obs-mip.fr/projets/narval/v1/} spectropolarimeter at T\'elescope Bernard Lyot (Observatoire Pic du Midi, France, \citealt{Auri2003}). 10 observations were made with the ESPaDOnS spectropolarimeter \citep{Espadons1,Espadons2} at the Canada France Hawaii Telescope (Mauna Kea Observatory, Hawaii) in 2017, which overlap the 2017 NARVAL observations. 

NARVAL and ESPaDOnS are twin spectropolarimeters, each with a spectral coverage of 3700 to 10480\,{\AA}  and a resolution of $\sim$65000 when used in polarimetric mode. The instruments both consist of a polarimeter mounted at the telescope's Cassegrain focus, which splits incoming circularly polarized light into two orthogonally-polarized beams that are then fibre-fed to the spectrograph and recorded on the detector. Each Stokes {\it{V}} observation is derived from a series of four sub-exposures, according to \citet{Donati1997}. The positions of the orthogonally-polarized beams are switched between sub-exposures, and then the Stokes {\it{V}} observation is determined by dividing sub-exposures with orthogonal polarization states. This removes spurious polarization signals to a first-order degree. Adding sub-exposures provides an unpolarized, Stokes {\it{I}} observation. A Null spectrum, which provides a measure of noise and a diagnosis of the reliability of the polarimetric measurement, can be obtained by dividing spectra with identical polarization states. 

NARVAL and ESPaDOnS observations were reduced and calibrated automatically at the TBL and CFHT using the software package {\sc{libre-esprit}} (based on {\sc{esprit}}, Echelle Spectra Reduction: an Interactive Tool, \citealt{Donati1997}). Bias and flat-field exposures taken on the observing night were used for the reduction, and a first-order wavelength calibration was carried out using a Thorium-Argon arc lamp exposure. This was further refined using telluric lines as RV references, as was first done by \citet{Donati2003_telluriclines}. 
Table \ref{tab:observation_summary} provides a summary of the spectropolarimetric observations organised into 4 epochs for magnetic field mapping, along with a summary of the HARPS, CARMENES and Kepler data used in this study. 

\begin{table*}
\caption{Summary of spectropolarimetric observations of GJ729 used for ZDI, and the HARPS, CARMENES and Kepler data used in this study. For each epoch, columns list the number of observations, the number of rotational cycles covered by the observations, the HJD of the first (HJD start) and last (HJD end) Stokes {\it{V}} observations, and the HJD corresponding to rotational cycle 0 for the epoch (close to the mid-point of the data, but an integer multiple of rotations from HJD=2454300). The final column shows the instruments that the observations were sourced with. 
}
\begin{tabular}{lcccccc}
\toprule
\multirow{2}{*}{Map} & \multirow{2}{*}{Observations} & \multirow{2}{*}{Rotations} & HJD start & HJD end & Mid HJD & \multirow{2}{*}{Instrument}  \\ 
  & & & (+2454300) & (+2454300) & (+2454300)  & \\ \midrule
2007 July-Aug.	&	9	&	5.59    &	11.49 & 27.41 & 	20.03	&		NARVAL \\
2008 Aug.		&	7	&	6.69    &	389.36 & 408.42 & 	398.81	&		NARVAL \\
2009 Aug.       &	18	&	3.54	&  756.38	&	766.46	& 760.51	&	 NARVAL \\
2017 Aug.	&	18	&	5.61	&  3669.40	&	3685.39	& 3674.01 &	 NARVAL + ESPaDOnS \\
2017 July-Sept.  & 62 & 32.25 & 3634.67 & 3727.51  &  3682.56 & HARPS \\
2017 July-Oct.  & 35 & 29.77 & 3648.49 & 3733.29 & 3691.10 &  CARMENES \\
2015 Oct.-Dec.  & 3449 & 28.56 & 3000.51 & 3082.85 & 3041.67 & Kepler\\
 \bottomrule
\end{tabular}
\label{tab:observation_summary}
\end{table*}

\subsection{K2 photometry}\label{sec:K2phot}

GJ 729 (EPIC 215632123) was observed by the Kepler space telescope in 30 min integrations during Campaign 7 of the K2 mission, from 2015 October 4 to 2016 April 22. The observations taken from the Mikulski Archive for Space Telescopes (MAST\footnote{https://mast.stsci.edu/portal/Mashup/Clients/Mast/Portal.html}) are shown in Figure \ref{fig:K2_timeseries}. The star's brightness shows clear rotational modulation with a period of $\sim2.848$d \citep{Ibanez2020} and a double dip pattern that prevails for the $\sim80$d of observations. 

%% file: ZDI.tex
\section{Magnetic field of GJ 729}\label{sec:magmapping}

\subsection{Least-Squares Deconvolution of spectral line profiles}\label{sec:LSD}

ZDI and PCA can characterize the large-scale stellar magnetic field based on rotationally modulated Zeeman signatures in Stokes {\it{V}} spectra. Zeeman signatures in individual Stokes {\it{V}} spectral lines are very small, so we used the technique of Least-squares deconvolution (LSD, \citealt{Donati1997}) to derive high signal-to-noise `average' spectral line profiles from the observations. LSD assumes that the spectral line asymmetries caused by magnetic activity are equal across all spectral lines. Therefore, the entire observed spectrum of an active star can be represented by a convolution of the LSD line profile, which captures changes in the line shape due to magnetic activity, and the spectrum of the star without magnetic activity impacts. 

Spectral lines for LSD were selected using a line mask from the Vienna Atomic Line Database (VALD, \citealt{Kupka2000}) covering 3500-11000\AA{} for a $T_{\mathrm{eff}}=3340\,K$, $\log g=5.23$, $\log\textrm{(M/H)}=-0.63$ star with microtubulent velocity of 1\,km\,s$^{-1}$. We excluded lines with depth less than 40 percent of the continuum and, as in \citet{Bellotti2023b, Bellotti2023}, excluded the wavelength bands of [6270,6320], [6555,6570], [6860,6970], [7160,7340], [7590,7700], [8130,8350], and [8950,9860] {\AA}, which are affected by telluric lines or in the vicinity of the H$\alpha$ line.  

\begin{figure}
    \centering
    \includegraphics[width=0.9\columnwidth]{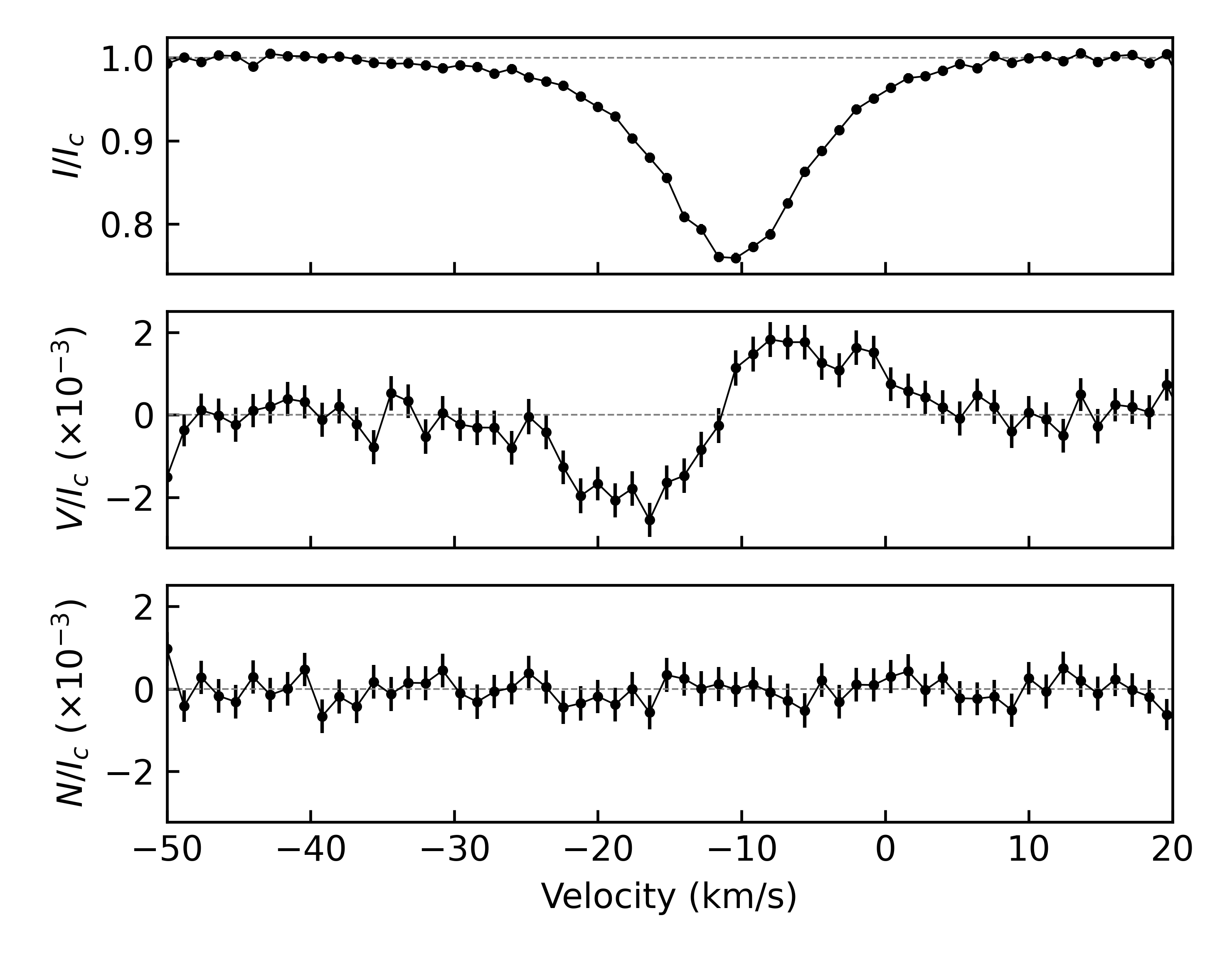}
    \caption{Example LSD profile for GJ 729. Axes (top to bottom) show the mean Stokes {\it{I}} profile, mean Stokes {\it{V}} profile and Null profile. The observation was taken on 2009 August 19.}
    \label{fig:LSDplot}
\end{figure}

LSD was performed using {\sc{ESpRIT}} \citep{Donati1997}, and using the line normalization parameters of wavelength 6700{\AA}, Lande factor 1.22 and central line depth 0.66 to bring the mean polarized and unpolarized line weights to roughly unity. These values are based on extending the table of \citet{marsden2014} to temperatures below 4000K, as was done in \citet{paper2}. An example of a continuum-normalized LSD line profile is shown in Figure \ref{fig:LSDplot},  and further details of the LSD technique can be found in \citet{Donati1997} and \citet{Kochukhov2010}.

\subsection{Surface-averaged longitudinal magnetic field and S index }

The longitudinal magnetic field, $B_l$ (G), is the line-of-sight magnetic field strength averaged over the visible stellar surface, and is measurable from the LSD line profiles. We measured $B_l$ from the NARVAL and ESPaDOnS LSD profiles using the process described in \citet{paper2}. We also measured the S-index, the ratio of the fluxes in the \ion{Ca}{ii} H and K lines to the continuum flux, from the intensity spectra, using the same method as in \citet{paper2}. The S-index and $B_l$ are shown as timeseries in Figure \ref{fig:ZDI_properties_timeseries}, along with the properties of the large scale magnetic field measured with ZDI. Both the S-index and $B_l$ show significant scatter at each epoch, such that variability is not significant between epochs. The $B_l$ is also impacted by flux cancellation due to the mixed polarity of the large-scale magnetic field (Section \ref{sec:field_evolution}).

\subsection{Zeeman-Doppler Imaging}\label{sec:ZDI} 

ZDI reconstructs the large-scale magnetic field by iteratively comparing model-generated LSD profiles with observed ones, aiming to fit the observations within a desired reduced-$\chi^2$ value while simultaneously producing a map of the large-scale magnetic field with maximum entropy (i.e. minimal image content) \citep{Donati2006}. The reconstructed magnetic map provides a lower-limit on the large-scale field strength \citep{Kochukhov2020}, but the recovered large-scale field structure is reliable.

In ZDI the model LSD profiles are constructed by representing the stellar surface as a grid of equal-area tiles, each characterized by a magnetic field vector with radial, azimuthal, and meridional components. The magnetic field for each tile is expressed as a series of spherical harmonics \citep{Donati2006}, with the maximum harmonic order, $l_{max}$, reflecting the complexity of the magnetic structure (e.g. $l=1$ for a dipolar structure, $l=2$ for quadrupolar structure, etc.). For each tile, the local magnetic field vector is projected onto the line of sight to compute the corresponding local polarised LSD profile. The global-scale signal is determined by integrating contributions from all tiles, which are scaled according to their projected area and to account for limb darkening, and Doppler shifted according to their line-of-sight projected rotational velocity. 

For this study, we used the ZDI implementation {\sc{ZDIpy}} \citep{Folsom2018a}, modified by \citet{Bellotti2023} to incorporate Unno-Rachkovsky’s solutions to polarised radiative transfer equations in a Milne-Eddington atmosphere \citep{Unno1956,Rachkovsky1967,LandiDeglinnocenti2004}. Additionally, the adaptation includes the filling factor formalism from \citet{Morin2008a}, where each surface tile is assumed to be uniformly covered by a fraction $f_{I}$ of magnetic regions, and a fraction $f_{V}$ of magnetic regions that produce the net polarization signature (i.e. a fraction $f_I - f_V$ accounts for small-scale, oppositely polarized features that cancel, producing no net polarization signature). The use of $f_V$ allowed us to fit to both the amplitude of the Stokes {\it{V}} profiles, which scales with the magnetic field strength (B), and the Zeeman splitting (the width of Zeeman signatures) which scales with $B/f_V$. 

\subsubsection{Stellar and line model parameters}
The stellar and LSD line-model parameters adopted for ZDI are listed in Table \ref{tab:lineParams}. As described in \citet{Bellotti2024}, the local LSD profiles were modeled using a Voigt profile, incorporating both a Gaussian component (representing thermal broadening) and a Lorentzian component (representing pressure broadening). Other model parameters include the ratio of the line to continuum opacity at line center ($\eta_0$, essentially the line strength), and the coefficient of the dependence of the source function on optical depth ($\beta$) in a Milne-Eddington atmosphere. Following \citet{Bellotti2024}, $\beta$ was calculated from the limb-darkening coefficient, assuming a linear limb-darkening law with a coefficient of 0.6813 \citep{Claret2011} for a star with $T_{\rm{eff}}=3500K$, $\log g=5.0$, Z=-0.6, and microturbulent velocity of 1 km\,s$^{-1}$. The Gaussian width, Lorentzian width and $\eta_0$ were optimized by iteratively varying these parameters and generating synthetic Stokes {\it{I}} line profiles. The parameters that minimized the reduced-$\chi^2$ fit to the Stokes {\it{I}} observations (without including a spot model) were selected. We adopted the same Gaussian and Lorentzian widths for all epochs, but used unique $\eta_0$ for each epoch to account for changes in the disk-integrated line strength associated with evolving magnetic activity.  The wavelength and Land\'e g-factor of the local line model matched those used in LSD (see section \ref{sec:LSD}). The maximum degree of harmonic expansion was set to $l_{max} = 8$ as the star's $v\sin i$ is much smaller than the intrinsic line width, such that only large-scale magnetic modes are recoverable. The adopted $l_{max} = 8$ represents a conservative upper limit rather than a restrictive choice, as higher-order modes would be weakly constrained and primarily fit noise.

\begin{table}
    \centering
    \caption{Stellar and line model parameters for ZDI}
    \begin{tabular}{lc}
    \toprule
    \textbf{Parameter} & \textbf{Value}  \\
    \midrule
        Wavelength   &   6700 {\AA} \\
        Gaussian width & 3.0 km s$^{-1}$\\
        Lorentzian width & 2.3 km s$^{-1}$ \\
        Limb-darkening coefficient & 0.6813 km s$^{-1}$ \\
        $v\sin i$  & 3.0 km s$^{-1}$\\
        Inclination angle &  55$\degr$  \\
        P$_{eq}$ & 2.848 d \\
        $l_{max}$ & 8 \\
        \bottomrule
    \end{tabular}
    \label{tab:lineParams}
\end{table}

The stellar $v\sin{i}$ and inclination angle were constrained using $\chi^2$ minimization, fitting to the Stokes {\it{V}} observations. The $\chi^2$ fit was minimized with $v\sin{i}$ between 2.7 and 3.3\,km\,s$^{-1}$, consistent with $3\pm1.5$\,km\,s$^{-1}$ from \citet{Reiners2018}. We adopted 3\,km\,s$^{-1}$. Based on this and using the published rotation period and stellar radius-mass relations, \citet{Reiners2018} inferred an inclination angle of $37\pm18$\,\degr. Conversely, \citet{Ibanez2020} adopted 90\degr. Significant rotational variability in Stokes {\it{V}} observations suggests a high inclination angle, and our ZDI models favored between 55\degr for our best data set (2017) and 70\degr. We adopted 55\degr, consistent with \citet{Reiners2018} within the uncertainties. We did also explore the effect of adopting differing inclination angles, see Section \ref{sec:field_evolution}.

We also used $\chi^2$ minimization to constrain the Stokes {\it{V}} filling factor, $f_V$. The reduced-$\chi^2$ was minimized with $f_V$ values of 0.15, 0.09, 0.12, and 0.18 for the  2007, 2008, 2009 and 2017 data sets respectively. For the final maps we adopted reduced-$\chi^2$ values of 0.9 for 2007, 2008, and 2009, and 1.3 for 2017. In these final maps, the reconstructed magnetic field was not significantly affected  by using using $f_V$ values up to 1.0.

\begin{figure}
    \centering
    \includegraphics[width=0.95\linewidth]{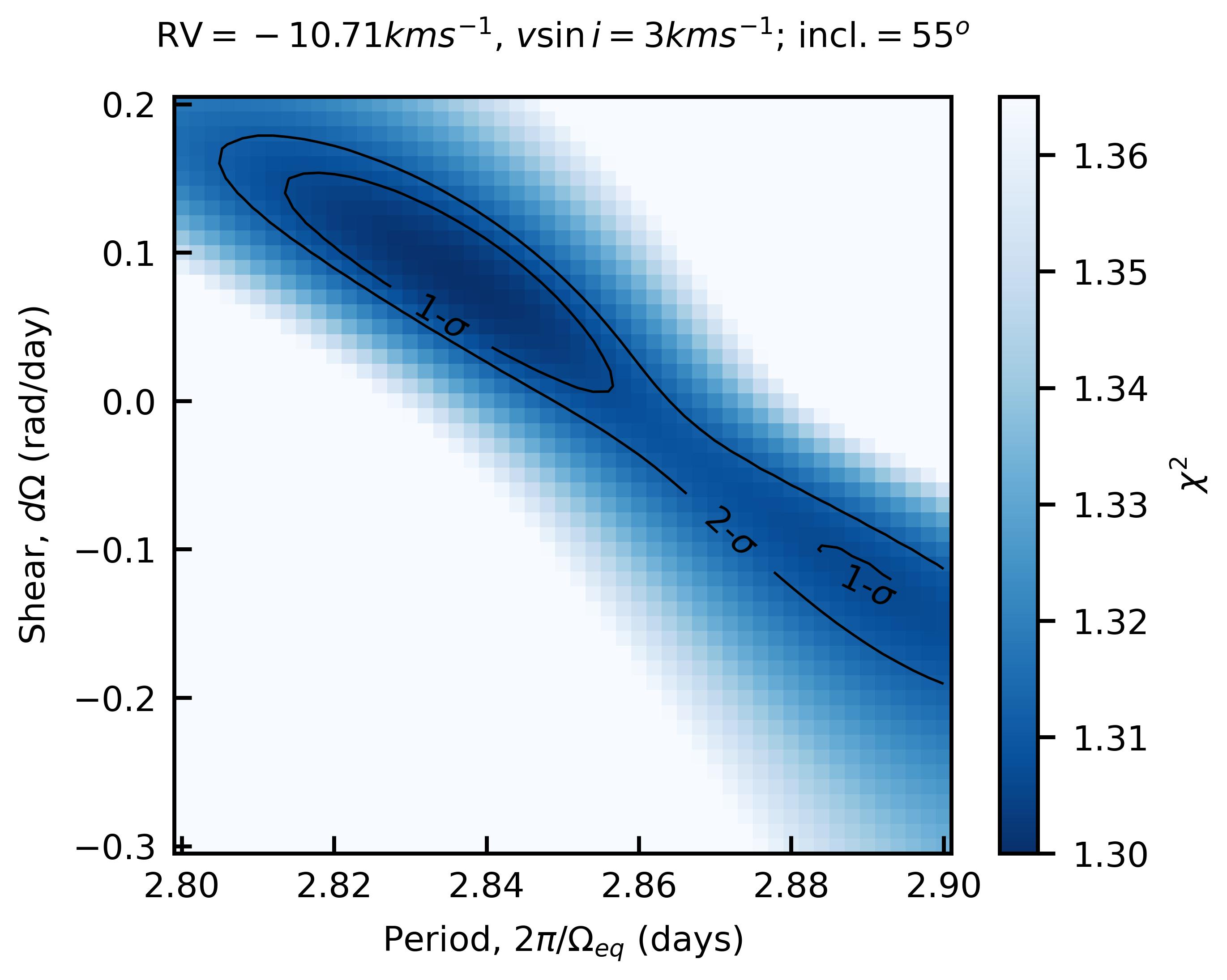}
    \caption{The reduced-$\chi^2$ landscape when fitting to the Stokes {\it{V}} profiles from 2017 with varying equatorial rotation period (d) and rotational shear (rad\,d$^{-1}$). The darkest blue region represents the best-fit combination of parameters, and 1$\sigma$ and 2$\sigma$ contours are also shown. }
    \label{fig:DiffRot}
\end{figure}

The rotational shear between the equator and poles (d$\Omega$) was estimated by assuming a solar-like differential rotation law:
\begin{equation}\label{DR_equation}
    \Omega (\theta) = \Omega_{\rm{eq}} - \rm{d}\Omega \sin^2\theta
\end{equation}
where $\Omega (\theta)$ is the angular rotation rate at a latitude $\theta$, and $\Omega_{\rm{eq}}$ is the angular rotation rate at the equator, all measured in $\rm{rad\,d^{-1}}$. In Figure \ref{fig:DiffRot} we show the reduced-$\chi^2$ landscape when fitting our 2017 data with varying equatorial rotation period and rotational shear. The 2017 data has the largest number of observations over 5.61 rotations, and is the best data set for tracing the differential rotation of magnetic features. The $\chi^2$ surface converges on both pro-grade and retrograde shear options, a known degeneracy when observations cover only a few stellar rotations. Given that the $1\sigma$ region encompasses a rotational shear of $0 \rm{\,rad\,d^{-1}}$, and \citet{Ibanez2020} found no differential rotation in their detailed spot modeling from K2 photometry, we have adopted solid-body rotation for our models. The fact that periodograms shown in Figures \ref{fig:Periodograms} and \ref{fig:K2_timeseries} do not show significant peaks at a spread of periods (that might correspond to the rotation of features at different latitudes), also supports solid-body rotation. We also tested the sensitivity of our ZDI results to variations in the rotation period and rotational sheer, as discussed in Section \ref{sec:field_evolution}. 

\subsubsection{Long-term evolution of the magnetic field from ZDI}\label{sec:field_evolution}

Key properties of the reconstructed field are summarized in Table \ref{tab:results}. The reconstructed large-scale magnetic field maps, the observed and model Stokes {\it{V}} LSD line profiles and a timeseries of selected magnetic field parameters from Table \ref{tab:results} are included in Appendix \ref{sec:ZDIresults_appendix}. As ZDI does not provide uncertainties for the reconstructed magnetic field parameters, we evaluated the sensitivity of the field parameters to changes in the adopted inclination angle, v$\sin$i, rotation period and rotational sheer using a process similar to \citet{Mengel2016}. We reconstructed the magnetic field for two parameter combinations corresponding to the extrema allowed by the $\chi^2$ landscape: (i) low inclination and low $v\sin i$ paired with a longer equatorial rotation period and no rotational shear $(37^\circ,\ 2.7~\mathrm{km\,s^{-1}},\ 2.86~\mathrm{d},\ d\Omega=0~\mathrm{rad\,d^{-1}})$, and (ii) high inclination and high $v\sin i$ paired with a shorter period and high shear $(70^\circ,\ 3.3~\mathrm{km\,s^{-1}},\ 2.814~\mathrm{d},\ d\Omega=0.15~\mathrm{rad\,d^{-1}})$. The maximum difference in the reconstructed magnetic field parameters between these realizations was adopted as the ‘variation bars’ shown in Table~\ref{tab:results} and Figure~\ref{fig:ZDI_properties_timeseries}.

\setlength{\tabcolsep}{3pt}
\begin{table*}
\caption{Large-scale magnetic field properties for GJ729 derived using ZDI. Columns (left to right) indicate the unsigned mean and maximum surface field strengths, the fraction of the total magnetic field stored in the poloidal component, the percentages of poloidal and toroidal field stored in dipolar ($l=1$), quadrupolar ($l=2$) and octopolar modes ($l=3$), the percentages of the total, poloidal and toroidal fields stored in axisymmetric modes and the strength and location of the negative pole of the radial component of the large-scale dipole. The upper and lower values shown for each measured parameter represent their sensitivity to variations in the $v\sin i$ (2.7 - 3.3 $km\,s^{-1}$), inclination angle (37 - 70 $\degr$), rotation period (2.814 - 2.860 d) and rotational sheer (0 - 0.15 rad d$^{-1}$).}
\begin{tabular}{lcccccccccccccccc}
\toprule
\textbf{Map} & 
{\textbf{$\rm{B_{mean}}$ (G)}} & {\textbf{$\rm{B_{max}}$ (G)}} & \multicolumn{4}{c}{\textbf{Poloidal}}              & \multicolumn{3}{c}{\textbf{Toroidal}}                        & \multicolumn{3}{c}{\textbf{Axisymmetric}}                             & \multicolumn{2}{c}{\textbf{Dipole radial -ve}}              \\ \cmidrule(lr){4-7} \cmidrule(lr){8-10} \cmidrule(lr){11-13} \cmidrule(lr){14-15}
                                   &   &    & \textbf{\% tot} & \textbf{$l=1$} & \textbf{$l=2$} & \textbf{$l=3$} & \textbf{$l=1$} & \textbf{$l=2$} & \textbf{$l=3$} & \textbf{\% tot} & \textbf{\% Pol} & \textbf{\% Tor} & \textbf{B (G)} & \textbf{Lat.} \\ \midrule
2007&  $82^{+14}_{-2}$  &  $279^{+122}_{-27}$  &  $69^{+2}_{-3}$  &  $55^{+6}_{-13}$  &  $33^{+3}_{-3}$  &  $10^{+5}_{-2}$  &  $78^{+1}_{-12}$  &  $20^{+10}_{-6}$  &  $1^{+4}_{-0}$  &  $43^{+2}_{-6}$  &  $20^{+1}_{-3}$  &  $93^{+0}_{-7}$  &  $-68^{+0}_{-10}$  &  $-20^{+3}_{-2}$   \\ [+1.5mm] 
2008&  $50^{+10}_{-2}$  &  $112^{+31}_{-10}$  &  $60^{+4}_{-5}$  &  $77^{+6}_{-8}$  &  $16^{+4}_{-4}$  &  $6^{+3}_{-2}$  &  $77^{+13}_{-16}$  &  $22^{+11}_{-13}$  &  $1^{+4}_{-0}$  &  $41^{+4}_{-3}$  &  $3^{+1}_{-1}$  &  $99^{+0}_{-2}$  &  $-61^{+0}_{-7}$  &  $-10^{+3}_{-2}$   \\ [+1.5mm] 
2009&  $50^{+21}_{-0}$  &  $179^{+144}_{-21}$  &  $90^{+0}_{-2}$  &  $38^{+10}_{-19}$  &  $41^{+3}_{-3}$  &  $17^{+13}_{-7}$  &  $73^{+0}_{-15}$  &  $23^{+10}_{-0}$  &  $3^{+4}_{-0}$  &  $17^{+5}_{-6}$  &  $10^{+6}_{-6}$  &  $79^{+0}_{-20}$  &  $-45^{+19}_{-11}$  &  $42^{+0}_{-7}$   \\ [+1.5mm] 
2017&  $145^{+26}_{-7}$  &  $541^{+346}_{-154}$  &  $49^{+1}_{-0}$  &  $37^{+9}_{-18}$  &  $21^{+1}_{-1}$  &  $30^{+5}_{-5}$  &  $72^{+12}_{-22}$  &  $24^{+10}_{-11}$  &  $3^{+8}_{-1}$  &  $48^{+2}_{-0}$  &  $3^{+2}_{-0}$  &  $91^{+2}_{-0}$  &  $-108^{+8}_{-5}$  &  $4^{+10}_{-7}$   \\ [+1.5mm]

\bottomrule
    \end{tabular}
    \label{tab:results}
\end{table*}

The average large-scale field strength from ZDI ranges from 50 to 145 G, and local field strengths reach 540 G. The global field varies between poloidal-dominated and roughly equal in toroidal-poloidal strength. The azimuthal field dominates at the surface for most epochs. The poloidal field is complex, with strong quadrupolar (up to 41 percent) and higher-order (up to 42 percent) modes, while the toroidal field is strongly axisymmetric and concentrated in dipolar and quadrupolar modes (above 96 percent). 

There is evidence for long-term field evolution but no clear magnetic cycle, which could indicate irregular dynamo behaviour or reflect incomplete phase coverage of the cycle. The mean field strength appears to weaken slightly from $82^{+14}_{-2}$ to $50^{+10}_{-2}$ G between 2007 and 2008, and then remains weak in 2009 at $50^{+21}_{-0}$ G. Notable structural changes occur between 2008 and 2009; the negative pole of the global dipole shifts from latitudes of $-20^{+3}_{-2}\degr$ in 2007 and $-10^{+3}_{-2}\degr$ in 2008, to the furthest position from the equator in 2009, at $42^{+0}_{-7}\degr$. This is accompanied by an apparent change in the polarity of the radial field across the visible pole in the magnetic field maps, and a strengthening of the large-scale poloidal field, which comprises 90 percent of the total magnetic field in 2009, compared to 49–69 percent in other epochs. By 2017, the negative pole of the large-scale dipole returns to a latitude of $4^{+10}_{-7}\degr$, and the mean field strength increases to nearly three times its 2009 value. Overall, the data indicate a weakening of the mean field and a poleward migration of the dipole between 2007 and 2009, followed by a recovery in field strength and equatorward dipole position by 2017. Further monitoring will be required to confirm whether this represents a cyclic or an irregular reconfiguration of the large-scale field. There is not any clear correlation between chromospheric activity and the large-scale magnetic field properties derived from ZDI, consistent with these diagnostics probing different components of the stellar magnetic field.

We caution that the Stokes {\it V} profiles from 2009 (Figure \ref{fig:StokesV}) show noticeably different shapes during the first two rotations, compared to the last rotation. The first 10 observations are all non-detections of the magnetic field, whereas 7 of the final 8 observations contain a clear Zeeman signature.  The differences in profiles could reflect rapid strengthening of the large-scale field, or a complex field configuration. Regardless, the full dataset can be fitted down to a reduced-$\chi^2$ of 0.90, and we opted not to split the epoch in two, as doing so would compromise phase coverage.

\input{PCA}

%% file: PCA.tex
\subsection{Principal Component Analysis}\label{sec:PCA}

Principal component analysis (PCA) was used by \citet{Lehmann2022} to probe the properties of large-scale stellar magnetic fields directly from timeseries Stokes {\it{V}} LSD profiles, without the detailed knowledge of stellar parameters that is a requirement of ZDI. PCA identifies key patterns in data using eigenvalues and eigenvectors, which represent respectively the main directions and strengths of variations in the data. Testing PCA on a range of stars is important for establishing it as a fast and reliable technique for large surveys, where ZDI cannot be applied as efficiently for many stars.

\begin{figure*}
    \centering
    \includegraphics[width=0.85\linewidth]{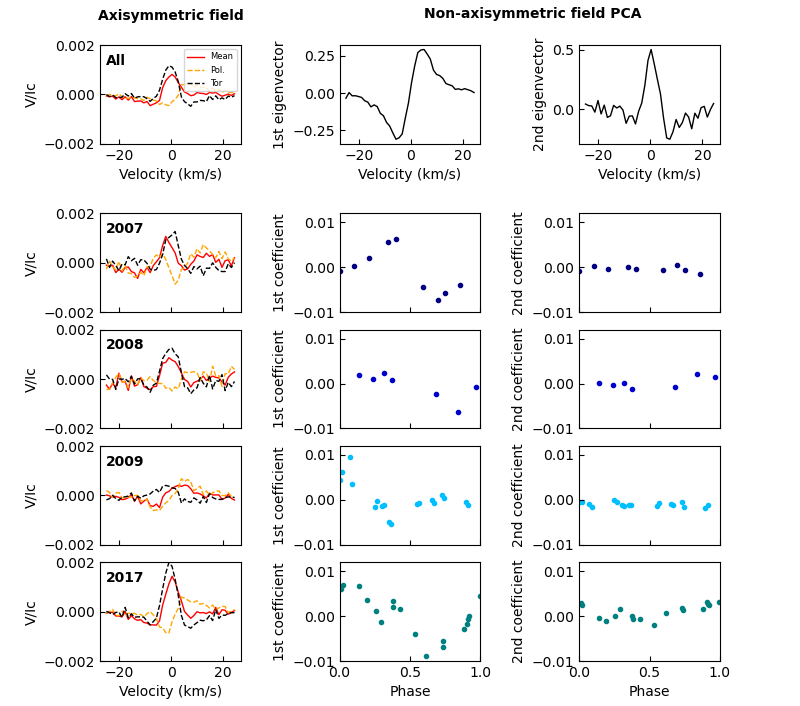}
    \caption{PCA analysis for GJ 729. Panels on the left indicate in red the time-averaged Stokes {\it{V}} profiles for all observations (top) and for the 2007, 2008, 2009 and 2017 observing seasons. The time-averaged profile is decomposed into the poloidal (asymmetric relative to the line center, orange dashed) and toroidal (symmetric, black dashed) components. The time-averaged Stokes {\it{V}} profiles indicate the structure of the axisymmetric component of the magnetic field. 
    Middle and right columns indicate the first and second eigenvectors for the mean-subtracted Stokes {\it{V}} observations, and their coefficients, for the (top to bottom) 2007, 2008, 2009 and 2017 observations. The mean-subtracted Stokes {\it{V}} profiles relate to the non-axisymmetric component of the magnetic field. }
    \label{fig:PCA}
\end{figure*}

Following \citet{Lehmann2022}, we computed the mean (or time-averaged) Stokes {\it{V}} profiles for our entire set of GJ 729 observations, and for each observational epoch individually, as shown in Figure \ref{fig:PCA}, left-hand column. The mean Stokes {\it{V}} profile describes the component of the large-scale field that is purely symmetric about the stellar rotation axis, for which the observed Zeeman signatures will be unchanged as the star rotates. The mean Stokes {\it{V}} profile is small compared to the mean-subtracted Stokes {\it{V}} profiles (see Figure \ref{fig:MeanStokesV2017} for example). This indicates that the field is predominantly non-axisymmetric. The axisymmetric component is at its weakest in 2009, and strongest in 2017. The mean line profile is predominantly symmetrical about the line-center, indicating a toroidal-dominated axisymmetric component. 

We further analyzed the non-axisymmetric component of the large-scale magnetic field by applying PCA on the mean-subtracted observations to determine a series of eigenvectors and corresponding eigenvalues. We subtracted the mean Stokes V profile and carried out PCA using the entire data set (rather than on an epoch-by-epoch basis)  to study the long-term variation of the coefficients of the eigenvectors.  The first two eigenvectors for the mean-subtracted Stokes {\it{V}} profiles are shown in Figure \ref{fig:PCA} at the top row, in the middle and right hand columns.  The first eigenvector accounts for 66 percent of the variance in the mean-subtracted Stokes {\it{V}} profiles. The second eigenvector shows a clear signal but it accounts for only 8 percent of the variance, and its coefficients are small. The remaining eigenvectors (not shown) contained only noise.  

The first eigenvector is asymmetric with respect to the line center, suggesting a multipolar field, although the coefficients indicate variations between dipolar and multipolar states. According to \citet{Lehmann2022}, the phased coefficients for the first eigenvector provide the most information about the complexity of the field. They found that a sinusoidal phase pattern is typical for a tilted dipolar field structure, and more complex phase variation is indicative of a multipolar structure. For GJ 729 the non-axisymmetric field appears to be multipolar in 2009 and 2017, but more strongly dipolar in 2007. The 2008 data has a large phase gap so the field structure is not easy to interpret. The 2008 phase variation appears to match the 2007 coefficients the closest, which might indicate a dipolar non-axisymmetric field. Interestingly, the phase variation seems to be reversed in 2009 relative to the other epochs, with the minimum appearing to phase with maxima for the other epochs (and vice versa) which could indicate the reversal of a quadrupolar or higher-order magnetic mode. 

\citet{Lehmann2022} also found that extrema in the phased coefficients traced individual, large magnetic features related to the poles of the low order multipoles. For 2007, the maximum coefficient at a phase of $\sim$0.4 and minimum at $\sim0.75$ appear to correspond to negative and positive radial field regions, respectively, in Figure \ref{fig:maps}. The 2008 coefficients also show a minimum that might correspond to a positive radial field spot in the ZDI maps. In 2009 and 2017, when the phase variation is complex, minima in the coefficients appear to correspond to regions of strong negative azimuthal field in the ZDI maps, at phases of $\sim0.25$ for 2009 and $\sim0.6$ for 2017.  

%% file: RV_Planet_Search.tex
\section{RV analysis}\label{sec:RVs}

\subsection{Stellar RVs and activity indicators}

We extracted RVs and activity indicators from the DRS-reduced HARPS spectra using {\sc{serval}} \citep{Zechmeister_SERVAL}.  {\sc{serval}} measures RVs by cross-correlating observations against a template spectrum derived from the observations themselves. First, each spectrum is cross-correlated with the highest signal-to-noise spectrum in the set to obtain initial RV estimates. Spectra are then co-added according to the initial RVs to create a template, and final RVs are determined by cross-correlation with the template spectrum. {\sc{serval}} excludes the ten bluest and ten reddest spectral orders, where the instrument efficiency decreases and telluric contamination can occur. The RVs are also corrected for instrumental drift and secular acceleration. The full set of RVs extracted from HARPS observations are shown in Figure \ref{RVobs_timeseries}, separated by color into archival observations (black) and the observations (red) used in this study. See \citet{Zechmeister_SERVAL} for a full description of the {\sc{serval}} products and their calculation. 

For CARMENES we downloaded RVs and activity indices from Data Release 1 of the Guaranteed Time Observations Data Archive \citep{Ribas2023}. We used the Average Corrected Velocity (AVC) provided in DR1, which was obtained from {\sc{serval}} RVs using nightly zero points to correct for instrumental systematics. AVCs were not calculated for 3 nights due to no instrumental drift value being available, so we have excluded these observations. The full set of CARMENES RVs is shown in Figure \ref{RVobs_timeseries}, with those used in our study shown in red.

The activity indices we study are the Differential Line Width (dLW), Chromatic Index (CRX), and H$\alpha$ spectral line index. The dLW measures changes in spectral line width relative to a reference spectrum, CRX is a measure of the RV variation with wavelength, and the H$\alpha$ index is the ratio of excess flux in the activity-sensitive H$\alpha$ line to continuum flux. We also computed the S-index from the HARPS observations using ACTIN2 (\citealt{daSilva2018}, see Figure \ref{fig:HARPS_Sindex}). CARMENES does not cover the wavelength range for the S index. 

We applied a $3\sigma$ clipping process to remove outliers, possibly related to flares, as was done by \citet{Zicher2022} for the flare star AU Mic. GJ 729 has a flare frequency of 0.5$d^{-1}$ and flare energies between $10^{32}$ and $10^{34}$ erg, which is two to four orders of magnitude more than the quiescent level \citep{Ibanez2020}. In total, we rejected 6 observations from HARPS and 2 from Carmenes. We also excluded the NARVAL and ESPaDOnS observations from our RV analysis as the uncertainties are at least an order of magnitude larger than the possible planetary modulations. 

Figure \ref{fig:RVs_filtered} shows the time series of RVs and activity indicators, with outliers removed. The activity indicators are also shown versus the RV. Only the CARMENES CRX shows a moderate anti-correlation with RV, consistent with the wavelength dependence of stellar activity at redder wavelengths. There is clear beating in the RV timeseries, especially the HARPS data, probably arising from the small frequency offset of $\sim0.05\rm{d}^{-1}$ between the stellar rotation frequency ($0.351\rm{d}^{-1}$) and the 1-2$f_{\rm{rot}}$ alias, which corresponds to a beat period of roughly 20d.

\subsection{Power spectra}

\begin{figure*}
    \centering
    \includegraphics[width=0.495\linewidth]{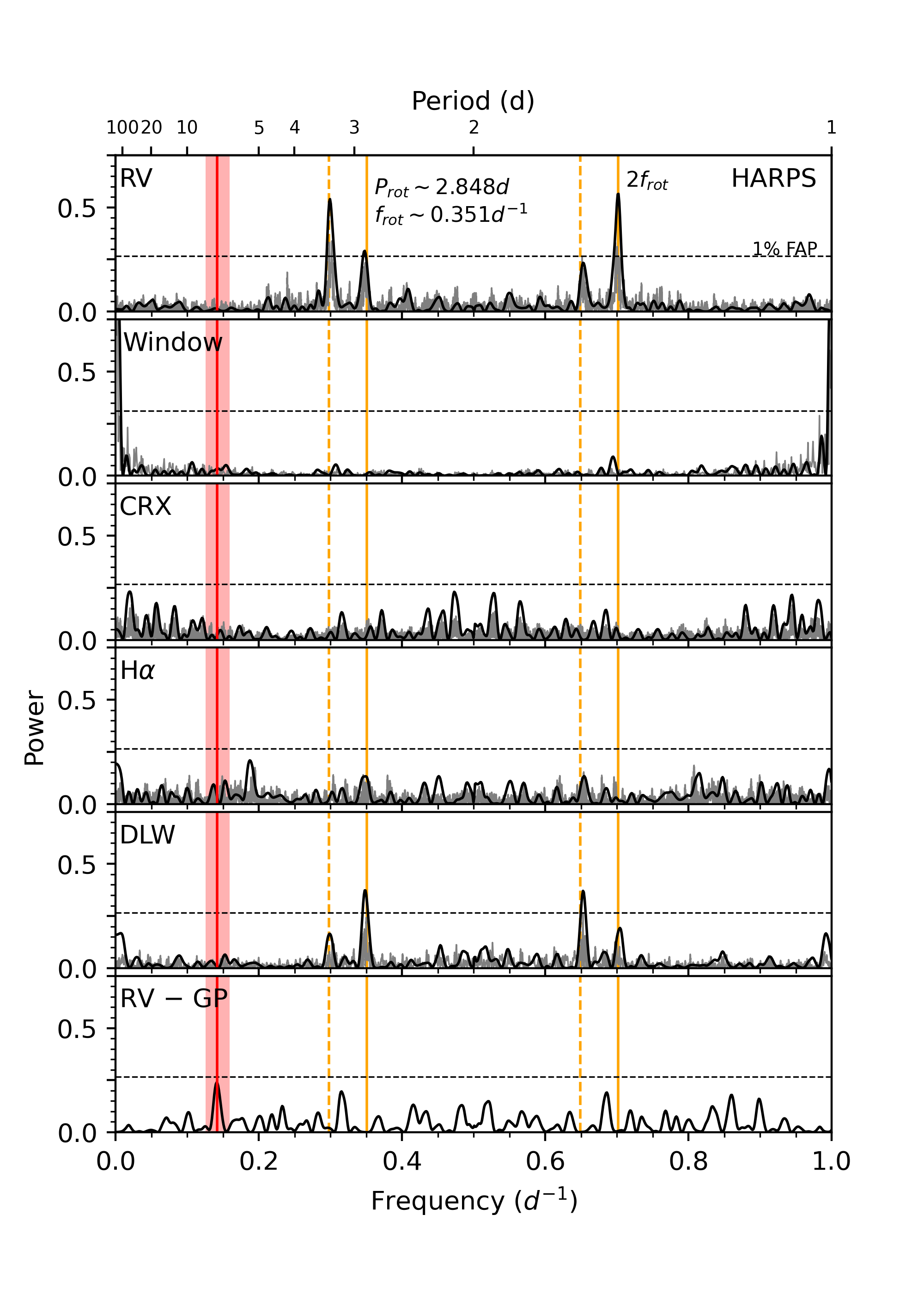}
    \includegraphics[width=0.495\linewidth]{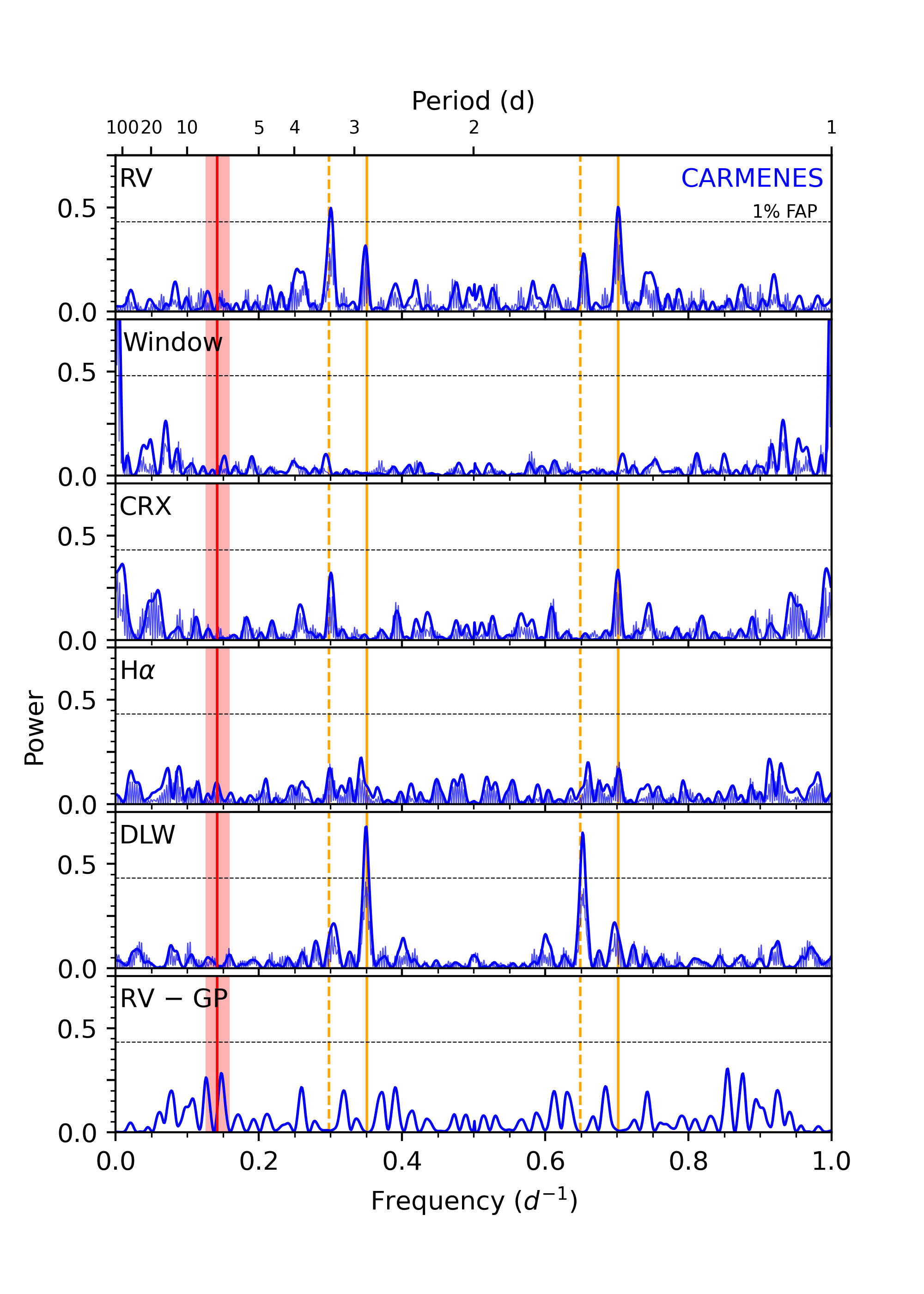}
    \caption{Top to bottom: Power spectra for the RV, window function, CRX, H$\alpha$, dLW and activity-subtracted RVs from HARPS (left) and CARMENES (right). The red vertical line and shaded region indicate a potential Keplerian signal with period of $7.11\pm0.84$d (see Section \ref{sec:pyorbit}). The dashed horizontal lines indicate 1 percent FAP.}
    \label{fig:Periodograms}
\end{figure*}

Power spectra for the RVs and activity indicators are shown in Figure \ref{fig:Periodograms}. The power spectra were derived using the Generalized Lomb-Scargle (GLS) periodogram \citep{Zechmeister2009}. Also shown in Figure \ref{fig:Periodograms} is the power spectrum for the Window function, which indicates patterns in the timing of observations. To assess signals beyond those induced by stellar activity, we also examined the residual RVs after subtracting a Gaussian Process (GP) activity model using {\sc{PyORBIT}} \citep{pyorbit}; see Section \ref{sec:activity_subtraction} for further details. Power spectra for the activity-subtracted RVs are shown in the bottom panel of Figure \ref{fig:Periodograms}.

\begin{figure*}
    \centering
    \includegraphics[width=0.8\linewidth]{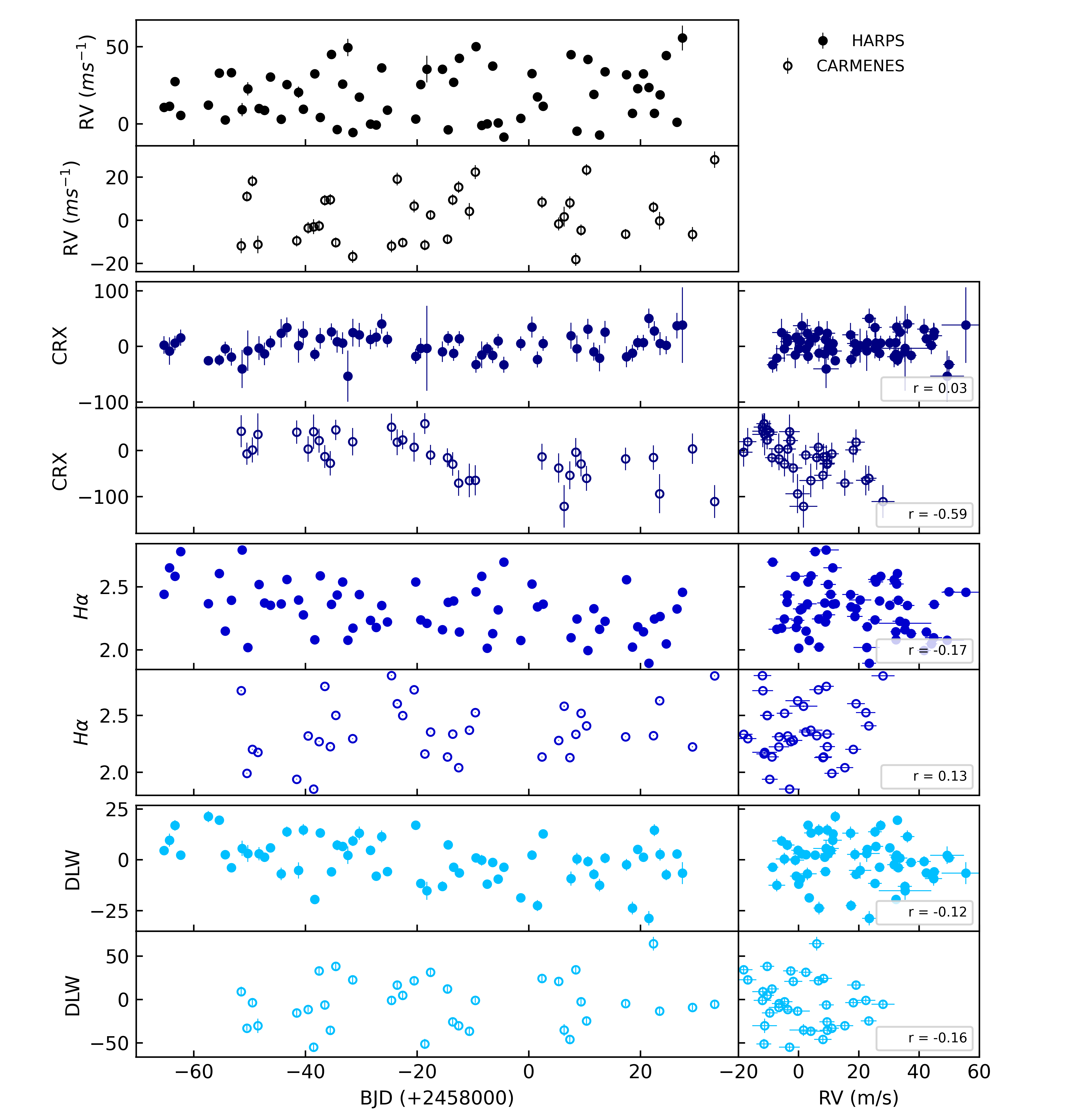}
    \caption{Left: Timeseries RV, CRX, H$\alpha$ and dLW from HARPS (filled circles) and CARMENES (open circles). Right: Activity versus RV for each of the CRX, H$\alpha$ and dLW, with markers the same as in the left hand column. Pearson correlation coefficients are also shown. }
    \label{fig:RVs_filtered}
\end{figure*}

For both the HARPS and CARMENES RVs, Figure \ref{fig:Periodograms} shows two significant peaks related to stellar activity at the rotation period and half-period (solid orange lines), which is expected for active stars \citep{Boisse2011}. The same signals are present in the K2 photometry (Appendix \ref{sec:K2phot}). The dLW is the only activity indicator that shows these peaks above a 1 percent False Alarm Probability (FAP), although there is no strong correlation between dLW and RV in Figure \ref{fig:RVs_filtered}. The RV power spectra also show 1d aliases of the rotation  period and half-period, indicated with orange dashed lines. 
Neither the CRX nor H$\alpha$ power spectra show any signals that are significant with respect to the 1 percent FAP. 

The activity-subtracted RVs show no periodicities with FAP < 1 percent. The strongest peaks occur at $\sim$7.14 d for the HARPS data and $\sim$6.90 d for CARMENES. These are further investigated in Section \ref{sec:activity_subtraction}. 

\subsection{RV modeling}\label{sec:activity_subtraction}

We tested three separate code packages to model the RV variations of GJ 729; {\sc{PyORBIT}} \citep{pyorbit}, {\sc{PYANETI}} \citep{Barragan_PYANETI}, and {\sc{Kima}} \citep{Kima}. In each case, we modeled activity as a Quasi-periodic Gaussian Process (GP), with a covariance matrix, $\lambda$, specifying the correlation between each pair of RV observations. The exact form of the quasi-periodic covariance matrix depends on the software package used. A circular Keplerian was used to model the RV variations induced by a single orbiting planet.  We fitted both GP-only and GP + Keplerian models to the data, evaluating their relative significance using Bayesian model comparison. 

\subsubsection{PyORBIT}\label{sec:pyorbit}
We jointly modeled the HARPS and CARMENES RVs using {\sc{PyORBIT}}, which supports the use of different GP hyperparameters for each instrument. As HARPS and CARMENES observe in different wavelength ranges, the amplitude of activity-induced RV variability can differ significantly, as shown by the different amplitudes of the RV and activity index modulations between the two instruments in Figure \ref{fig:RVs_filtered}. {\sc{PyORBIT}} uses nested sampling through \texttt{PyMultiNest} \citep{Pymultinest2014} for model parameterization and to estimate the Bayesian evidence, allowing for robust model comparison. 

To model stellar activity we used a quasi-periodic GP kernel from \citet{Rajpaul2015}, implemented in {\sc{PyORBIT}} through {\sc{george}} \citep{GEORGE2015}.  The kernel has the form
\begin{equation}
\gamma (t_i, t_j) = A^2  \exp{ \left \{-\frac{\sin^2{[\pi(t_i - t_j)/ P]}}{2 O^2} - \frac{(t_i-t_j)^2}{2 P_\mathrm{dec}^2} \right \} }
\end{equation}
where $A$ is the amplitude, $P$ is the recurrence timescale ($\sim \rm{P_{rot}}$), $O$ the coherence timescale or inverse harmonic complexity (indicating how complex the signal is within one period, <1 representing high frequency variations and >>1 being low frequency) and $P_{\rm{dec}}$ the decay time scale of the active regions. The uncertainty of each RV measurement $\sigma_i$ and an additional white noise term $\sigma_j$ for each instrument were added in quadrature to the diagonal of the covariance matrix.  An RV offset was also included for each observing instrument ($\rm{RV_{HARPS}}$ and $\rm{RV_{CARMENES}}$).

Priors and posterior parameters for GP-only and GP + Keplerian models are provided in Table \ref{tab:GP_results_Pyorbit}. We allowed for an instrument-dependent GP amplitude, while all other GP hyperparameters were shared between instruments. Since the stellar rotation period is well constrained we adopted tight priors for the GP recurrence timescale ($P$). For the amplitudes, coherence timescale and decay timescale we adopted broad, uninformative priors. 

\begin{figure}
    \centering
    \includegraphics[width=\linewidth]{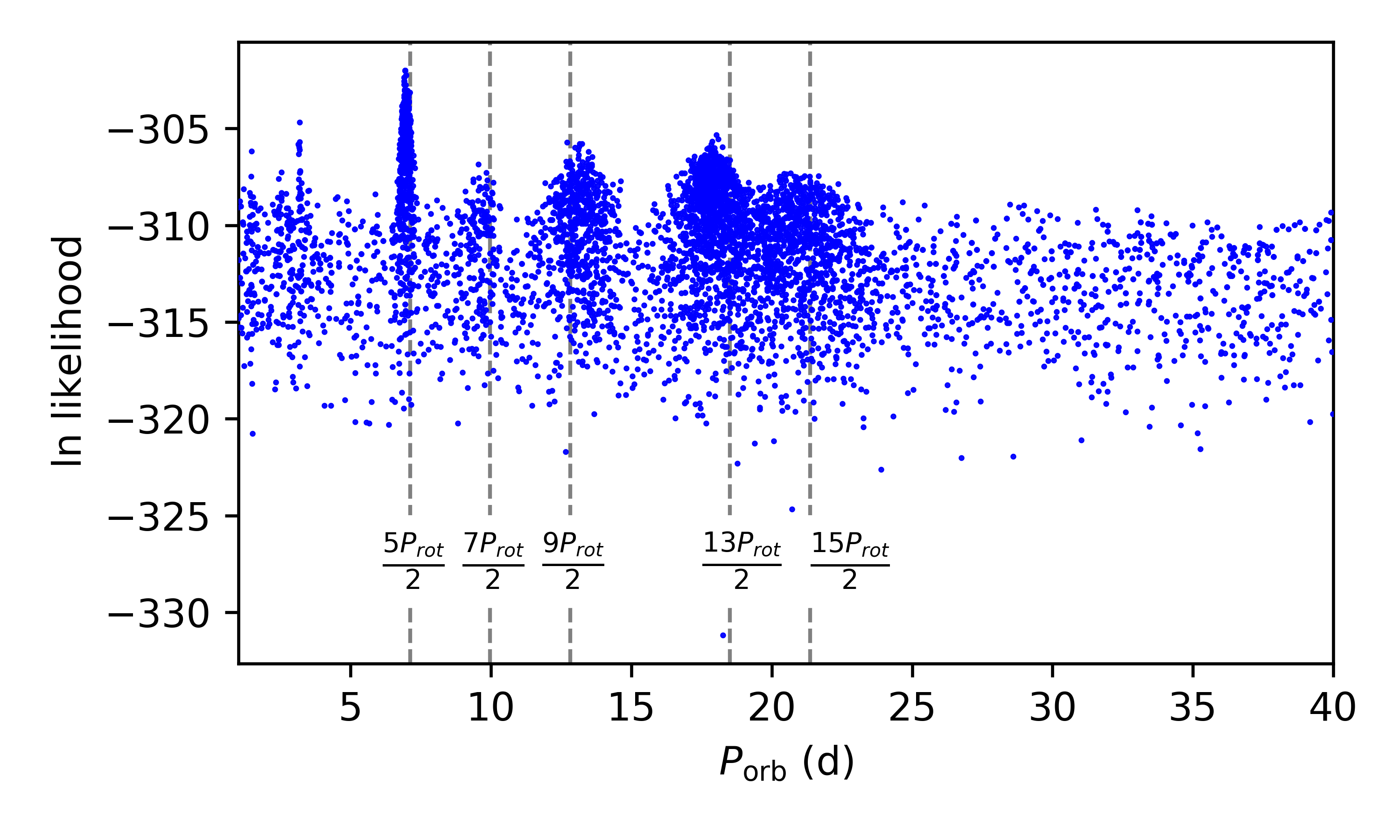}
    \caption{Log-likelihood versus orbital period (d) for samples from a GP + Keplerian model fit to the HARPS and CARMENES data with {\sc{PyORBIT}}. Samples concentrate at odd multiples of $\frac{P_{rot}}{2}$ and the most likely orbital period is $\sim$7d.  } 
    \label{fig:PversusLnProb}
\end{figure}

For the GP + Keplerian model, we adopted broad, uninformative priors for the planet amplitude. We initially searched a broad parameter space for the planet orbital period, using priors of $\mathcal{U}$[1, 40]\,d. We found the posterior distribution consisted of a series of signals occurring at odd multiples of half the rotation period, as shown in Figure \ref{fig:PversusLnProb}. We further tested the stability of these signals under different modeling assumptions, including a quasiperiodic + cosine GP kernel \citep{Perger2021} which includes a cosine component to improve the fit to the first harmonic of the stellar rotation period. In the case of GJ 729, the first harmonic is stronger than the rotation period itself, as shown in Figure \ref{fig:Periodograms}. We also tested a fit to an extended time series including HARPS and CARMENES data within 1000 d of the main dataset, and tested a joint fit to the RV and dLW activity indicator. Unlike {\sc{Pyaneti}} (Section \ref{sec:pyaneti}), which jointly models RV and activity indicators as functions of both a GP and its time-derivative, the {\sc{PyORBIT}} implementation does not include the derivative term. The posterior log-likelihood versus orbital period for each of these tests is included in Appendix \ref{sec:RVtests}. A $\sim7$\,d signal consistently emerged as the highest-likelihood solution. Motivated by these tests, we restricted the orbital period prior to $\mathcal{U}$[4, 10]\,d to assess the significance of candidate signals within this region of parameter space. 

We fitted models to the RVs using the nested sampling Monte Carlo routine {\sc{PyMultiNest}}. The parameter space was explored by 500 live points. {\sc{PyMultiNest}} monitors convergence by estimating the maximum possible contribution to the Bayesian evidence from the remaining prior parameter space at each iteration. For our models, sampling was considered converged when this contribution fell below 0.5.

\setlength{\tabcolsep}{10pt}
\begin{table*}
    \centering
    \caption{Priors and posterior parameters for RV models fitted using {\sc{PyORBIT}}. The mean and standard deviation of the posterior distributions are shown, with the maximum-a-priori values shown in square brackets.  For the priors in the second column, $\mathcal{U}$ represents Uniform priors, with bracketed values indicating the minimum and maximum. }
    \begin{tabular}{lccc}
    \toprule
     \bf{Parameter} & \bf{Prior}& \bf{GP} & \bf{GP + Keplerian}\\ \midrule
    $A_{\rm{HARPS}}$ (m\,s$^{-1}$)  & $\mathcal{U}$ [1,50]   & $24.35\pm6.68$ [17.32] & $24.75\pm6.41$ [21.84]\\
    
    $A_{\rm{CARMENES}}$ (m\,s$^{-1}$)  & $\mathcal{U}$ [1,50]   & $16.25\pm4.14$ [13.94] & $15.38\pm3.66$ [14.50] \\    
    
    $P_{\rm{dec}}$ (d) & $\mathcal{U}$ [1,90]  & $71.04\pm11.03$ [64.75] & $67.92\pm11.72$ [62.62] \\
    
    $P$  (d)   & $\mathcal{U}$ [2.84,2.86]    & $2.848\pm0.002$ [2.847] & $2.848\pm0.002$ [2.847] \\
    
    $O$ (d)     & $\mathcal{U}$ [0.1,2]     & $0.38\pm0.06$ [0.35] & $0.39\pm0.05$ [0.39] \\
    
    $\rm{RV_{HARPS}}$ (m\,s$^{-1}$) &  $\mathcal{U}$ [$\rm{RV_{min}}$, $\rm{RV_{max}}$]   & $21.79\pm12.51$ [21.55] &  $20.48\pm11.91$ [11.17] \\
    
    $\rm{\sigma_{HARPS}}$  (m\,s$^{-1}$) & $\mathcal{U}$ [0, 5] &$2.39\pm0.56$ [2.27] & $1.81\pm0.68$ [1.70] \\
    
    $\rm{RV_{CARMENES}}$ (m\,s$^{-1}$) &  $\mathcal{U}$ [$\rm{RV_{min}}$, $\rm{RV_{max}}$]   &  $1.08\pm8.77$ [-2.34] & $1.55\pm7.86$ [1.31] \\
    
    $\rm{\sigma_{CARMENES}}$  (m\,s$^{-1}$) & $\mathcal{U}$ [0, 5] & $0.98\pm0.74$ [0.48] & $0.87\pm0.63$ [0.40]\\
    
    \midrule
    $P_{\rm{orb}}$ (d) & $\mathcal{U}$ [4, 10] & & $7.11\pm0.84$ [6.99] \\
    $K$ (m\,s$^{-1}$)  & $\mathcal{U}$ [0.1, 5]  & & $1.61\pm0.64$ [1.88]\\
    $T_c$ (BJD - 2458000) & $\mathcal{U}$ [0, 10] & & $5.45\pm2.36$ [6.58] \\
    \midrule
    Marginal evidence, $\ln Z$ &&  $-342$ & $-341$   \\
    Bayes Factor, $\Delta \ln Z$ && & $+1$  \\
    \bottomrule
    \end{tabular}
    \label{tab:GP_results_Pyorbit}
\end{table*}

\begin{figure*}
    \centering
    \includegraphics[width=0.9\linewidth]{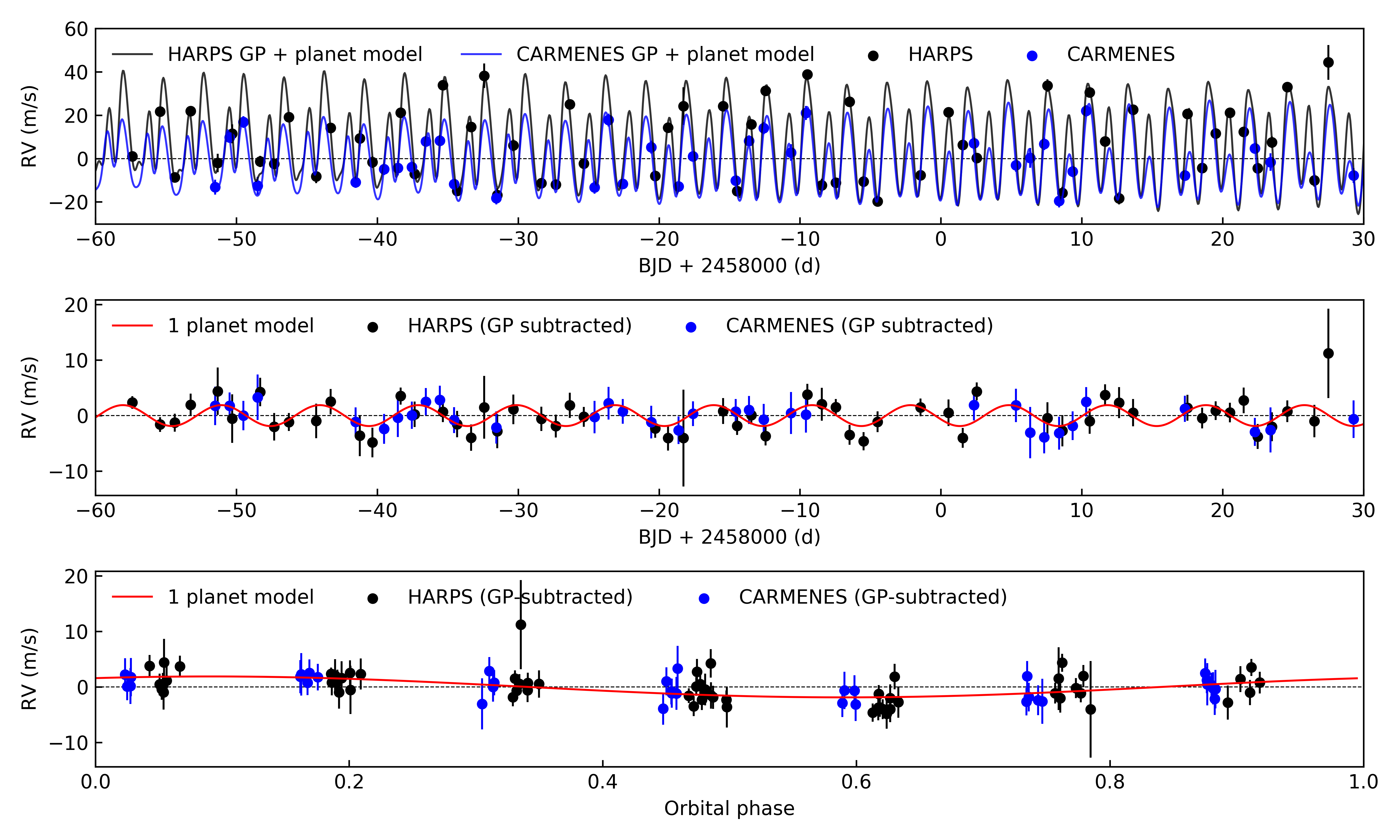}
    \caption{Top: HARPS (black) and CARMENES (blue) RVs are plotted against the maximum likelihood GP + Keplerian models fitted using {\sc{PyORBIT}}. The GP amplitude is different for the HARPS and CARMENES fits, but the other GP hyperparameters and the Keplerian model are fitted jointly. \\
    Middle: GP-subtracted RV residuals from HARPS and CARMENES are shown against the maximum likelihood planet model (red line).\\
    Bottom: Same as the middle plot, but phase folded using the maximum likelihood planet orbital phase from Table \ref{tab:GP_results_Pyorbit}.}
    \label{fig:RV_fit}
\end{figure*}

The GP-only and GP + Keplerian models yielded statistically comparable fits. The GP + Keplerian model had a very low Bayes Factor ($\Delta \ln Z$) of 1 when compared against the GP-only model. According to \citet{Trotta2008} a more complex model can be accepted when $\Delta \ln Z > 5$.  The best-fitting GP + Keplerian model had an orbital period of $6.99$\,d and amplitude of $1.88\,\rm{m\,s}^{-1}$, which would equate to a $\sim1.5M_{\oplus}$ planet orbiting at $\sim0.04$\,au, closer to the star than the habitable zone of 0.064-0.127 au estimated by \citet{Hardegree_Ullman_2023}. Figure \ref{fig:RV_fit} shows the maximum-a-prioiri GP + Keplerian model against the observed RVs, and the most-likely Keplerian fit to the GP-subtracted residuals. A corner plot showing the parameter correlations and posterior distributions for the GP + Keplerian model is also given in Figure \ref{fig:corner_PYORBIT}. The posterior GP parameters were not significantly different between the GP-only and GP + Keplerian models.  We also note that when we further constrained the priors on the planet orbital period to $\mathcal{U}$ [6.9,7.2], the maximum-likelihood and posterior distributions of model parameters were not significantly changed, and the marginal evidence increased to -338, with a Bayes Factor of 4 when compared to a GP-only model. 

Since the candidate signal is close to $5P_{\rm rot}/2 \simeq 7.12$ d, stellar activity cannot be excluded as the source of the $\sim7$ d signal. It is possible that the subtraction of the activity model redistributes power to this subharmonic of rotation. However, it is not obvious why such a feature would be consistently recovered across multiple modeling approaches employing different kernels, as shown in Appendix \ref{sec:RVtests} and in the following sections.

\subsubsection{\sc{PYANETI}}\label{sec:pyaneti}

We carried out further tests using {\sc{PYANETI}}, which is capable of jointly modeling activity-induced RV shifts ($V(t)$) and activity indicators ($\alpha(t)$) as functions of the same GP-drawn function, $\lambda(t)$, and (unlike {\sc{PyORBIT}}) its time-derivative, $\dot{\lambda}(t)$, according to 

\begin{equation}
    V (t) = A_0 \lambda(t) + A_1 \dot{\lambda}(t) + V_{o}\end{equation}\begin{equation}
    \alpha (t) =B_0 \lambda(t) + B_1 \dot{\lambda}(t) + \alpha_{o}
\end{equation}

where the free parameters $A_0$, $A_1$, $B_0$ and $B_1$ relate the individual timeseries with the GP function and its first-derivative. Instrumental offsets are accounted for by $V_o$ and $\alpha_0$ for each observing instrument. We chose to model simultaneously the RV and dLW, which is the only activity indicator to show significant periodicities at the stellar rotation period. The use of the time-derivative of the GP is important because in RVs the activity signal depends on the spot size and contrast, but also changes significantly with time as bright and dark spots move across the disk. A spot near the limb produces the largest spectral line asymmetry and RV shift, with the sign of the RV shift depending on which limb the spot is on. A spot near the disk center produces minimal RV variation.  Meanwhile, the dLW, which measures changes in line width relative to a reference spectrum, is mostly dependent on the size and contrast of active regions. dLW is less sensitive to the location of the spot on the disk, and the dLW should not switch sign as a spot transits from one hemisphere to the other. For this reason, we set $B_1$ to 0 such that there is no dependence of dLW on the GP time-derivative.  

For $\lambda(t)$ we used a quasi-periodic covariance function, as implemented by {\sc{PYANETI}}: 

\begin{equation}\label{eq:GP}
    \lambda_{ij}=\rm{exp} \left[ - \frac{(t_i - t_j)^2}{2\lambda_e^2} - \frac{\sin^2(\frac{\pi(t_i - t_j)}{P_{GP}})}{2\lambda_P^2} \right] 
\end{equation}

where the hyper-parameter $P_{GP}$ is the recurrence time-scale of features, $\lambda_P$ is the inverse harmonic complexity, and $\lambda_e$ is the evolution timescale, or the typical lifetime of stellar surface features. An instrumental offset was included, and the uncertainty in each data point, $\sigma_i$, and an extra uncorrelated noise parameter, $\sigma_s$, were added in quadrature and applied to the diagonal of the covariance matrix. 

{\sc{PYANETI}} can simultaneously model multiple instruments but only with shared GP hyper-parameters. After some testing we chose to model HARPS and CARMENES data separately with {\sc{PYANETI}}, as the instruments observe in different wavelengths and recover activity differently.  Modeling both instruments with shared GP hyper-parameters leaves residual power at the stellar rotation period and half-period.

{\sc{PYANETI}} uses the MCMC package {\sc{emcee}} \citep{ForemanMackey_emcee} for model parameterization. We used 100 Markov chains to explore the parameter space, checking for convergence every 5000 steps. Chains were taken as converged when the Gelman criterion R was < 1.02 for all the sampled parameters \citep{gelman2013bayesian, Barragan_PYANETI_1, Barragan_PYANETI}. We then used the last 5000 iterations of converged chains and thinned them by a factor of 10 to create the posterior distributions. We note that the model evidences from {\sc{PYANETI}} are parametric estimates.  {\sc{PyORBIT}} and {\sc{Kima}} use Nested Sampling to integrate over the entire parameter space, which is more computationally expensive but considered to be a more robust method for estimating model evidences. As such, the evidences from {\sc{PYANETI}} can not be directly compared against those from {\sc{PyORBIT}} and {\sc{Kima}}. The adopted priors, inferred parameters and parametric model evidences are shown in Table \ref{tab:GP_results}. 

\renewcommand{\arraystretch}{1.3}
\renewcommand{\tabcolsep}{10pt}
\begin{table*}
    \centering
    \caption{Priors and posterior parameters for GP and GP + Keplerian models fitted to the RV and dLW using {\sc{PYANETI}}. For the priors in the second column, $\mathcal{U}$ and $\mathcal{M}$ represent Uniform and Modified Jeffrey's (see \citealt{Gregory2007}) priors, with bracketed values indicating the minimum and maximum. }
    \begin{tabular}{lccccc}
    \toprule
     \bf{Parameter} & \bf{Prior}& \multicolumn{2}{c}{\textbf{HARPS}} & \multicolumn{2}{c}{\textbf{CARMENES}}\\
     \cmidrule(lr){3-4} \cmidrule(lr){5-6} 
     &  &  \bf{GP} & \bf{GP + Keplerian} & \bf{GP} & \bf{GP + Keplerian} \\ \midrule
    $A_0$ (m\,s$^{-1}$)  & $\mathcal{U}$ [0, 50]    & $3.41_{-1.90}^{+3.12}$ & $4.46^{+3.24}_{-2.00}$ & $2.57_{-1.86}^{+3.68}$ & $2.26_{-1.59}^{+2.97}$ \\
    $A_1$ (m\,s$^{-1}$) & $\mathcal{U}$ [0, 50]  & $10.14_{-2.99}^{+5.27}$ & $10.03^{+4.72}_{-2.84}$ & $10.75_{-3.57}^{+4.81}$ & $10.62_{-3.57}^{+4.72}$  \\
    $B_0$    (m\,s$^{-1}$)   & $\mathcal{U}$ [-100, 100]     & $16.66_{-4.92}^{+8.26}$ & $16.38^{+7.59}_{-4.63}$   & $57.33_{-19.52}^{+25.43}$ & $56.49_{-19.69}^{+25.15}$  \\
    $P_{\rm{GP}}$ (d)     & $\mathcal{U}$ [2.84, 2.86]     & $2.8485^{+0.0033}_{-0.0031}$ & $2.8518^{+0.0033}_{-0.0035}$  & $2.8490_{-0.0039}^{+0.0044}$   & $2.8501^{+0.0045}_{-0.0044}$   \\
    $\lambda_{P}$   & $\mathcal{U}$ [0.1, 2]      & $0.587_{-0.079}^{+0.104}$ & $0.600^{+0.096}_{-0.079}$  & $0.831_{-0.175}^{+0.210}$ & $0.840_{-0.185}^{+0.215}$ \\
    $\lambda_{e}$ (d)  & $\mathcal{U}$ [1, 90]      & $55.99_{-12.83}^{+18.45}$ & $44.90^{+11.29}_{-7.66}$ & $76.58_{-15.29}^{+9.64}$ & $77.27_{-14.93}^{+9.26}$\\
    $\rm{RV_{inst}}$ (m\,s$^{-1}$) &  $\mathcal{U}$ [-50,50]    & $18.65_{-1.80}^{+2.77}$ & $19.19^{+3.02}_{-2.38}$ & $0.59_{-2.05}^{+2.11}$ & $0.66_{-1.70}^{+1.84}$ \\
    $\rm{\sigma_{RV, inst}}$  (m\,s$^{-1}$) & $\mathcal{M}$ [1,1000]  & $2.02_{-0.65}^{+0.63}$ & $0.68^{+0.79}_{-0.51}$ & $0.77_{-0.57}^{+1.00}$ & $0.68_{-0.52}^{+0.95}$  \\
    $\rm{dLW_{inst}}$  (m\,s$^{-1}$)   &  $\mathcal{U}$ [-50,50] & $3.82_{-9.99}^{+10.64}$ &$3.00^{+9.89}_{-9.57}$ & $-7.14_{-40.08}^{+41.41}$ & $-6.96_{-38.42}^{+37.80}$ \\
    $\rm{\sigma_{dLW, inst}}$  (m\,s$^{-1}$)  & $\mathcal{M}$ [1,1000]  & $5.17_{-0.57}^{+0.67}$ & $5.20^{+0.70}_{-0.58}$ & $14.57_{-2.00}^{+2.47}$ & $14.44_{-1.98}^{+2.46}$ \\
    \midrule
    $P_{\rm{orb}}$ (d) & $\mathcal{U}$ [4, 10] & & $7.10_{-0.09}^{+0.09}$ & &  $6.84_{-3.05}^{+0.14}$  \\
    $K$ (m\,s$^{-1}$)  & $\mathcal{U}$ [0, 5]  & & $2.11_{-0.53}^{+0.52}$ & & $1.84_{-1.19}^{+1.11}$ \\
    $T_{\rm{peri}} (BJD - 2458000)$ & & & $-1.78^{+0.02}_{-0.02}$ & & $-1.71^{+0.76}_{-0.03}$ \\
    M ($M_{\oplus}$) &  & & $2.05_{-0.52}^{+0.51}$ & & $1.69_{-1.12}^{+1.06}$ \\
    \midrule
    Marginal evidence, $\ln Z$ & & 461 & 469 & 229 & 233   \\
    Bayes Factor, $\Delta \ln Z$ & & & +8 & & +4  \\
    \bottomrule
    \end{tabular}
    \label{tab:GP_results}
\end{table*}

We found a GP + Keplerian model to be preferred over the GP-only model for the HARPS data, with a Bayes Factor of 8. However, we note that the evidence computed from {\sc{PYANETI}} is a parametric estimate only, and is not derived by integrating over the entire parameter space. Therefore, the evidences shown in Table \ref{tab:GP_results} may not be as robust as those from Table  \ref{tab:GP_results_Pyorbit}. For CARMENES data the GP + Keplerian model offered a statistically equivalent fit to the GP-only model, with a Bayes factor of 4. 

The posterior distributions of Keplerian parameters from {\sc{PYANETI}} are consistent with those obtained from the two-instrument fit with {\sc{PyORBIT}}. HARPS data support a $2.05^{+0.51}_{-0.52}\rm{M}_{\oplus}$ Keplerian with an orbital period of $7.10\pm0.09$ d, whereas CARMENES data support a slightly lower-amplitude signal at $1.69^{+1.06}_{-1.02}\rm{M}_{\oplus}$ with a shorter orbital period of $6.88^{+0.14}_{-3.05}$ d. The amplitudes and periods of the signals are consistent within their respective uncertainties, and align roughly in phase, as indicated by the fitted time of periapsis, $T_{\rm{peri}}$. The small difference in the recovered periods could be attributed to undersampling in the CARMENES observations.

There are some differences in the GP properties between HARPS and CARMENES, but the recurrence timescale is consistent across the data. HARPS showed a larger activity-induced RV amplitude compared to CARMENES, and more rapid activity evolution timescale. This is to be expected, given that the instruments observe in different wavelengths, and highlights the importance of modeling different instruments with different GP hyperparameters in the case of high magnetic activity.  

We also tested fits to RVs only. Results indicated a small decrease in the evidence for a Keplerian in the CARMENES data, but no change for HARPS data.  This suggests that joint modeling of RVs and activity indicators may be most beneficial when the RV time series is more sparsely sampled. We also tested joint fits to RVs with CRX and H$\alpha$, but the models did not adequately fit the observed variability in those indicators nor lead to an improved description of the RVs.

\subsubsection{Kima}

Unlike {\sc{PYANETI}} and {\sc{PyORBIT}}, which require the number of Keplerians to be fixed, {\sc{Kima}} can treat the number of Keplerians as a free parameter, enabling flexible and rigorous model comparison. {\sc{Kima}} uses Diffusive Nested Sampling \citep{Brewer2009} to explore the parameter space and compute the fully marginalized posterior probability (the Bayesian evidence). {\sc{Kima}} does not support joint fitting to RVs and activity indicators, nor the use of a time-derivative GP term.  Like with {\sc{PYANETI}}, we modeled HARPS and CARMENES RVs separately as {\sc{Kima}} requires shared GP hyperparameters when combining instruments.

To model activity we adopted a quasi-periodic GP kernel from \citet{Haywood2014}, given by: 
\begin{equation}\label{eq:GP}
    \lambda_{ij}=\eta_1^2 \cdot \rm{exp} \left[ - \frac{(t_i - t_j)^2}{2\eta_2^2} - \frac{2\sin^2(\frac{\pi(t_i - t_j)}{\eta_3})}{\eta_4^2} \right] + (\sigma_i^2 + j^2)\delta_{ij}
\end{equation}
where the hyper-parameter $\eta_1$ is the RV semi-amplitude related to activity, $\eta_2$ is the decay parameter, $\eta_3$ is the recurrence time-scale and $\eta_4$ relates to the harmonic complexity. Note that the harmonic complexity as implemented in {\sc{Kima}} is related to that in {\sc{PYANETI}} and {\sc{PyORBIT}} as $\eta_4 = 2 \lambda_e = 2O$.

We tested models with the number of Keplerians as a free parameter between 0 and 1. Priors and the maximum likelihood parameters are shown in Table \ref{tab:GP_results_Kima}. Note that {\sc{Kima}} samples over both 0- and 1-planet models, such that the posterior distributions for Keplerian parameters are mixtures of two distinct hypotheses. As a result, we do not report uncertainties for the parameters shown in Table \ref{tab:GP_results_Kima}, since they would not represent the fully conditional posterior. Instead, these results show that when a model including a Keplerian is favored, the recovered period and amplitude are consistent with our other modeling approaches.  
With {\sc{Kima}}, a Keplerian with a 7.09 d orbital period and semi-amplitude of 2.24 m s$^{-1}$ was fitted to the HARPS data, with a probability (over a GP-only model) of 52 percent. For the CARMENES data, a GP-only model was preferred, with the GP + Keplerian model having a probability of only 31 percent.  We also tested tighter priors on the orbital period ($\mathcal{U}$[6.5, 7.5] d) and found the probability of a single Keplerian increased to 81 percent for HARPS and 38 percent for CARMENES. Importantly, changing the priors did not significantly impact on the most-likely orbital parameters. 

\renewcommand{\tabcolsep}{5pt}
\begin{table}
    \centering
    \caption{Priors and maximum likelihood parameters for RV models fitted using {\sc{Kima}}. For the priors in the second column, $\mathcal{U}$ represents Uniform priors, with bracketed values indicating the minimum and maximum. }
    \begin{tabular}{lccc}
    \toprule
     \bf{Parameter} & \bf{Prior}& \bf{HARPS} & \bf{CARMENES}\\ \midrule
    $\eta_1$ (m\,s$^{-1}$)  & $\mathcal{U}$ [0,50]   & 17.56 &  13.60 \\
    $\eta_2$ (d) & $\mathcal{U}$ [0,90]  & 40.35 & 85.02  \\
    $\eta_3$ (d)   & $\mathcal{U}$ [2.84,2.86]    & 2.849 & 2.851 \\
    $\eta_4$ (d)     & $\mathcal{U}$ [0.1,2]     & 0.71 & 0.77 \\
    $\rm{RV_{inst}}$ (m\,s$^{-1}$) &  $\mathcal{U}$ [$\rm{RV_{min}}$, $\rm{RV_{max}}$]   & 18.03 & 0.99 \\
    $\rm{\sigma_{inst}}$  (m\,s$^{-1}$) & $\mathcal{U}$ [0, 5] & 0.0017 & 0.0005 \\
    \midrule
    $P_{orb}$ (d) & $\mathcal{U}$ [4,10] & 7.09 & 6.81 \\
    $K$ (m\,s$^{-1}$)  & $\mathcal{U}$ [0, 5]  & 2.24 & 2.64 \\
    \midrule
    P($N_P=1$) && 52\% & 31\% \\
    \bottomrule
    \end{tabular}
    \label{tab:GP_results_Kima}
\end{table}


\subsection{Training the GP using photometry}

With $\sim80$d of 30 min-cadence K2 photometry available (2015 October to 2016 April), we tested training the GP on photometry to provide tighter constraints on the RV activity model. Using photometry to inform RV GPs has been effective even when the datasets are not contemporaneous \citep{Beard2025}.  We fitted a GP-only model to the photometry (binned by 0.1 days to reduce computational cost) and then used the resulting posterior parameter distributions as priors for a GP-only fit to the RVs. However, the RV fit tended toward the edges of the priors derived from the photometry. In particular, the RV variability displayed higher complexity and a longer decay timescale than inferred from the photometry. Given the strong activity variability in GJ 729, the time gap between datasets likely means that the star is exhibiting different activity levels, spot configurations, and phases of its activity cycle. Therefore, we chose not to constrain our final RV analysis using parameters derived from the K2 photometry.

\subsection{BGLS periodogram analysis of GP-subtracted RVs}

Figure \ref{fig:Periodograms} (bottom panels) shows GLS periodograms for the residual RVs after subtracting the GP-only model, fitted jointly to HARPS and CARMENS observations using {\sc{PyORBIT}}. A $\sim7$d periodicity is present within both the HARPS and CARMENES power spectra, but has FAP > 1 percent. 

\begin{figure}
    \centering
    \includegraphics[width=0.85\linewidth]{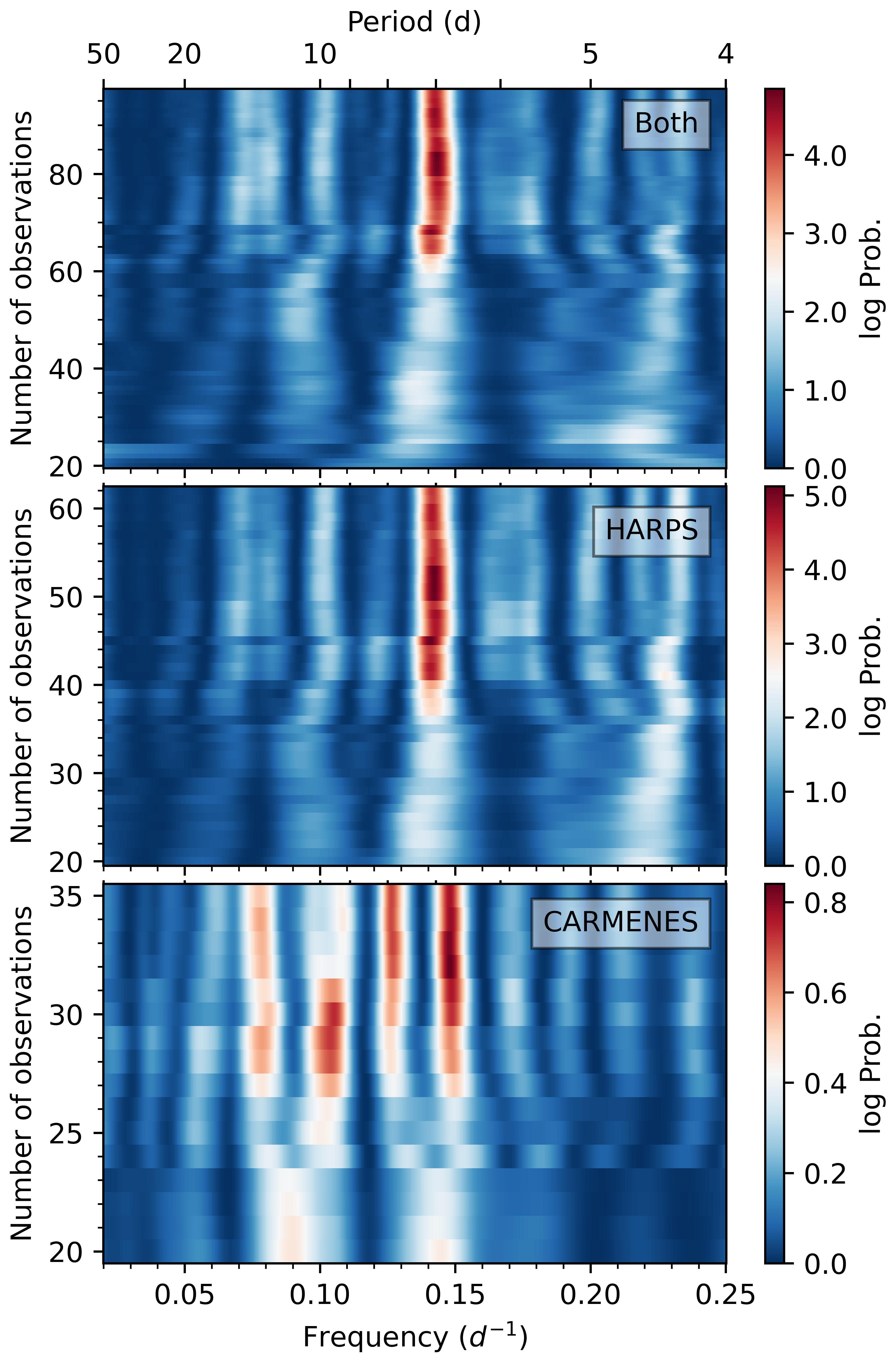}
    \caption{Stacked Bayesian GLS periodograms for the residual RVs following subtraction of the GP-only activity model fitted jointly to HARPS and CARMENES data using {\sc{PyORBIT}}. }
    \label{fig:BGLS}
\end{figure}

We further analysed the residuals of the GP-only model using a stacked Bayesian Generalised Lomb-Scargle (BGLS) periodogram in Figure \ref{fig:BGLS}.  The BGLS periodogram \citep{Mortier2015} estimates the Bayesian probability that a sinusoidal signal of a given period is present in the data. Following the approach of \citet{Mortier2017}, we computed the BGLS periodogram for the first $n$ data points, added the next observation, recomputed the periodogram, and repeated this process until the end of the timeseries. To track the evolution of potential signals, we normalised each periodogram by its minimum probability, allowing us to compare how the signal strength evolves over time. 

As discussed by \citet{Mortier2017}, a long-lived periodic signal should exhibit a steadily increasing BGLS probability as more data are added. In contrast, activity-driven signals that are typically quasi-periodic, with evolving amplitude and phase, may show fluctuating BGLS power as more data are added. For GJ\,729, however, activity-induced variations can remain coherent for at least 80 d (Figures \ref{fig:RVs_filtered} and \ref{fig:K2_timeseries}), consistent with the presence of long-lived active regions. The CARMENES data display a $\sim$6.9\,d signal that grows in probability as observations are added. For the HARPS and combined data, the BGLS probability of a $\sim$7.1\,d signal increases initially, before decreasing slightly following the final HARPS observations. The stacked BGLS analysis does not rule out residual stellar activity as the source of the signal.

\subsection{Planet detection thresholds}

\begin{figure}
    \centering
    \includegraphics[width=\columnwidth]{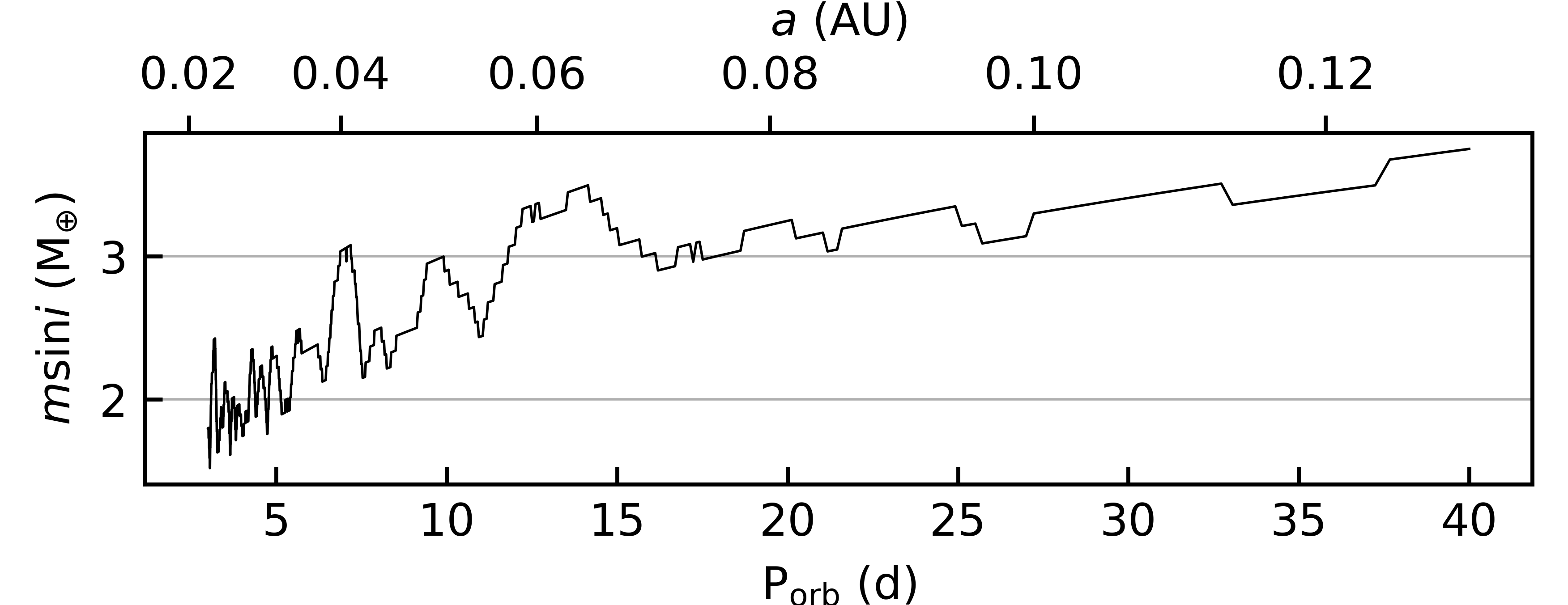}
    \caption{Minimum detectable mass versus orbital period for exoplanets in a circular orbit around GJ 729. The detection limit shown by the black line is the 99.9 percent confidence limit for detection in the GP-subtracted HARPS and CARMENES observations. We include horizontal grid lines at different masses for clarity. }
    \label{fig:Detection_limits}
\end{figure}

Finally, we derived period–mass limits above which we can exclude planets around GJ 729 in our activity-subtracted HARPS and CARMENES RVs using the same process as in \cite{Brown2023_V889her}. We estimated the 0.1 percent FAP threshold for signal detection by generating 10,000 shuffled data sets (times preserved, RVs randomised), and taking the periodogram power exceeding 99.9 percent of maxima. We then injected simulated circular Keplerian signals into residuals from the {\sc{PyORBIT}} GP-only fit, at trial orbital periods between 3 and 40 d and 12 equally spaced phases between 0 and 1, starting with an amplitude $K=2\,\rm{m\,s^{-1}}$.  We increased the RV amplitude in 0.1\,$\rm{m\,s^{-1}}$ increments until the periodogram power for the relevant period exceeded the FAP. A planet was considered detectable when this occurred for all 12 orbital phases. The resulting detection limits are shown in Figure \ref{fig:Detection_limits}. 

The detection-limit curve shows several spikes at periods where coherent features remain in the activity-subtracted RVs, particularly in the CARMENES observations (see Figure \ref{fig:BGLS}). At these periods, injected Keplerian signals can interfere destructively (or constructively) with the existing variability, so larger amplitudes are required for all phases to cross the false alarm threshold. Away from these features, the limits are smoother and reflect our typical sensitivity to planets. From this curve, we can rule out planets with minimum masses of about $3M_{\oplus}$ in the habitable zone of 0.064-0.127\,au \citep{Hardegree_Ullman_2023}.  

%% file: Discussion.tex
\section{Discussion and conclusions}\label{sec:D&C}

This paper presents a combined study of the long-term magnetic field evolution and search for exoplanets around the nearby, active M dwarf GJ 729. 

\subsection{Magnetic field}
We applied ZDI to series of Stokes {\it{V}} observations from NARVAL and ESPaDOnS, covering four epochs between 2007 and 2017. ZDI revealed a 50 to 145 G mean large-scale field strength. The field configuration was shown to be multipolar, varying from poloidal dominated (up to 90 percent) to a near-equal poloidal-toroidal configuration (51 percent toroidal) on a timescale of months-to-years. The field properties and their evolution implies GJ 729 belongs to the class of mid-to-late M dwarfs with weaker, non-axisymmetric and rapidly evolving magnetic fields \citep{Morin2010,Morin2011}. 

We further investigated the large-scale magnetic field geometry using PCA, obtaining results consistent with ZDI. PCA confirmed a predominantly non-axisymmetric large-scale field, with the weaker axisymmetric field dominated by the toroidal component. PCA and ZDI showed that the field was at its weakest, most poloidal and non-axisymmetric in 2009. ZDI results suggested that the negative pole of the dipolar axis moved from near the equator in 2008 to the mid-latitudes in 2009, and returned to the equator by 2017. Meanwhile, PCA results could indicate a temporary change in polarity of a higher-order multipole in 2009. A polarity change could be a feature of a magnetic cycle or an irregular reconfiguration of the large-scale field. 

GJ 729 is thought to undergo potentially multiple activity cycles, with a short $\sim$3.5–7 year cycle possibly modulated by a longer-term cycle \citep{Ibanez2020,Irving_2023}. We could not confirm any chromospheric activity cycles based on the new data presented here, and the sparse temporal sampling prevents us from determining whether the magnetic field evolution is cyclic or instead reflects irregular evolution of the magnetic field. \citet{Ibanez2020} showed a potential increase in S-index between 2007 and 2009, while our data show no significant change in S-index during this period and a decrease in large scale field strength from ZDI. The relationship between chromospheric activity and the large-scale magnetic field is not yet clear, motivating further monitoring to characterise this behaviour.

\subsection{Exoplanet search}

We searched for planetary companions using high-cadence, overlapping sets of precise RV measurements from HARPS and CARMENES. We modeled data from both instruments together, and separately, using three RV modeling packages; {\sc{PyORBIT}}, {\sc{Pyaneti}} and {\sc{Kima}}. In each case, we modeled activity as a quasi-periodic GP, with and without a circular Keplerian, and compared the models using Bayesian evidence estimates. We consistently recovered a signal that could relate to a $\sim1.5-2M_{\oplus}$ planet in a $\sim7$d orbit. However, the GP + Keplerian model did not offer a statistically significant improvement over a GP-only model. As \citet{ForemanMackey_emcee} discuss, flexible GPs can absorb low-amplitude periodic signals, which makes distinguishing such signals from activity particularly difficult. In the case of GJ 729, the {signal} has an amplitude of $\sim1.9\,\rm{m\,s}^{-1}$, a small fraction of the activity amplitude, measured to be $17 - 24\,\rm{m\,s}^{-1}$ for HARPS data. 

Near-7\,d signals of similar amplitude were recovered independently in the HARPS and CARMENES datasets and across multiple modeling approaches, and the signal was not clearly present in activity indicators. BGLS periodogram analyses of the activity-subtracted RVs showed that the BGLS probability of the $\sim$7\,d signal increased monotonically for the CARMENES data as observations were added, but not for the HARPS data. The recovered period is close to five times the stellar rotation half-period, and may be part of a ladder of signals at odd multiples of the rotation half-period. While activity-induced RV variability is typically associated with $P_{\rm rot}$ and its harmonics in frequency space \citep{Boisse2011}, our results suggest that, in highly active stars, residual structure can also appear at higher-order multiples of rotation and its harmonics, following GP modeling. Consistent with this, the $\sim7$\,d signal only becomes apparent after subtraction of the dominant activity signal, indicating that it may arise from structure not fully captured by the GP model.

To validate or formally exclude a planet we would require further long-term, nightly spectroscopic observations, ideally taken at near-infrared wavelengths where activity effects are reduced. Given the proximity of GJ 729 to Earth, at $\sim9.8$\,ly, confirmation of a planet would make it a compelling target for future detailed characterisation.